\documentclass[12pt]{article}

\usepackage{graphicx,amsmath,amssymb,array,calc}
\usepackage{bm}
\usepackage{times}

\newcommand{\bea}{\begin{eqnarray}}
\newcommand{\eea}{\end{eqnarray}}
\newcommand{\be}{\begin{equation}}
\newcommand{\ee}{\end{equation}}
\newcommand{\nn}{\nonumber}
\usepackage{graphicx,fullpage}
\usepackage{amsmath,amssymb,array,calc,amsfonts}
\newcommand\frakfamily{\usefont{U}{yfrak}{m}{n}}
\DeclareTextFontCommand{\textfrak}{\frakfamily}

\renewcommand\appendix{\par
\setcounter{section}{0}%
\setcounter{subsection}{0}%
\setcounter{table}{0}
\setcounter{figure}{0}
\setcounter{equation}{0}
\gdef\thetable{\Alph{table}}
\gdef\thefigure{\Alph{figure}}
\gdef\theequation{\Alph{section}-\arabic{equation}}
\gdef\thesection{\Alph{section}}
\setcounter{section}{0}}

\begin{document}

\begin{titlepage}


%

\vskip 4cm

 \begin{center}
{\bf \large Color kinematic symmetric (BCJ) numerators 
in a light-like gauge}
\end{center}
\vskip 1cm

\centerline{\large  Diana Vaman\textsuperscript{*}, York-Peng Yao\textsuperscript{$\dagger$} {}\footnote{E-mail addresses: dv3h@virginia.edu, yyao@umich.edu}
}

\vskip .5cm
\centerline{\it \textsuperscript{*} Department of Physics, The University of Virginia}
\centerline{\it Charlottesville, VA 22904, USA}
\centerline{\it \textsuperscript{$\dagger$}  Department of Physics, The University of Michigan}
\centerline{\it Ann Arbor, MI 48109, USA}
\vspace{1cm}

\begin{abstract}

Color-ordered tree level scattering amplitudes in Yang-Mills theories can be written as a sum over terms which display the various propagator poles of Feynman diagrams. The numerators in these expressions which are obtained by straightforward application of Feynman rules are not satisfying any particular relations, typically. However, by reshuffling terms, it is known that one can arrive at a set of numerators which satisfy the same Jacobi identity as the corresponding color factors. By extending previous work by us we show how this can be systematically accomplished within a Lagrangian framework. We construct an effective Lagrangian which yields tree-level color-kinematic symmetric numerators in Yang-Mills theories in a light-like gauge at five-points.  The five-point effective Lagrangian is  non-local and it is zero by Jacobi identity. The numerators obtained from it respect the original pole structure of the color-ordered amplitude. 
 We discuss how this procedure can be systematically extended to higher order. 
\end{abstract}

\end{titlepage}

\section{Introduction}

In field theory, a very intriguing result is the Kawai-Lewellen-Tye (KLT) relation \cite{KLT}, which 
connects gauge theories with gravity.  A particular version of the KLT relations was given by Bern, Dennen, Huang and Kiermaier in \cite{GGsq}. Following \cite{GGsq}, let us give a brief 
summary about it in the language of Lagrangian field theory.  We symbolically write the tree color-dressed
n gluon amplitude as
\be
A^{(n)}=\sum_i {c_in_i\over (\Pi s_j)_i}, \label{1.1}
\ee
where $c_i$ are color factors, $n_i$ are  
numerators made of momenta and polarization vectors, 
and $(\Pi s_j)$ are appropriate products of inverse propagators,
all constructed according to a well defined set of Feynman rules
once a gauge choice is made. Bern, Carrasco and Johansson \cite{BCJ}  stated that for channels
which satisfy the Jacobi identity
\be
c_i+c_j+c_k=0,  \label{1.2}
\ee
one can reshuffle terms and obtain a new set of numerators so that they seem  
to have been constructed completely through some effective three point
vertices and thus also satisfy
\be
\bar n_i+\bar n_j+\bar n_k=0.  \label{1.3} 
\ee
The relations (\ref{1.2}) and (\ref{1.3}) have been called color-kinematics duality and the numerators which have this property have been called BCJ or color-kinematic symmetric.  

Together with the antisymmetry of the effective three vertices,
(\ref{1.2}) leads to a reduction in the number of independent color
coefficients to $(n-2)!$, and via (\ref{1.3}) to the same number of
independent numerators.  Thus, naively one may also conclude
that there are $(n-2)!$  independent color-ordered 
amplitudes.  Any of such a set is called a Kleiss-Kuijf basis \cite{KK}.  We can
form a column vector for a set of the independent numerators $|\bar N\rangle$
and another  column vector for the set of color-ordered amplitudes in the chosen Kleiss-Kuijf basis $|A\rangle$. They are related by
\be
|A\rangle =M|\bar N\rangle , \label{1.4}
\ee
where the elements of the propagator matrix $M$ are made of sums
of products of propagators.  An appropriate set of independent
color coefficients will form a row vector $\langle C|$, which will yield the 
color-dressed $n$ particle amplitude 
\be
A^{(n)}=\langle C|M|\bar N\rangle .  \label{1.5}
\ee
In \cite{VY3} we pointed out that in fact there are $(n-3)(n-3)!$ degrees of 
arbitrariness in changing the elements in $|\bar N\rangle $, which will yield
the same $|A\rangle$.  This freedom in writing up the BCJ numerators was called the generalized gauge transformations in \cite{VY3}\footnote{In \cite{GGsq} the authors exploited this freedom in their proof of the relationship between gravity and gauge theory amplitudes.}. 
The underlying reason for this is that there are $(n-3)(n-3)!$ eigenvectors
with zero eigenvalue for $M$\footnote{Recent work by Cachazo et al. expressed the entries of what we called the propagator matrix $M$ in \cite{VY3} as partial amplitudes of a double-copy scalar theory with cubic interactions and wrote them in terms solutions to the so-called scattering equations \cite{Cachazo:2013iea}.} and therefore one can add to $|\bar N\rangle$ this number of arbitrary functions, each multiplied  to one of the zero eigenvectors.  Clearly, it has no effect on $|A\rangle.$  Seen through this, the true number of independent elements in $|A\rangle$ is in fact only $(n-3)!$. 

It is important that we should be in our possession a set of dual symmetric
numerators, because the KLT relation, as expressed by Bern et al in \cite{GGsq}, in the present context is a statement that up to
coupling constants, the tree level $n$ graviton amplitude is given by
\be
A^{(n)}_{gr}=\langle \tilde N|M|\bar N\rangle,\label{1.6}
\ee
in which $\tilde n_i$ can be numerators due to a different gauge 
theory or not, which satisfy the color-kinematic duality relations
\be
\tilde c_i+\tilde c_j+\tilde c_k=0, \label{1.7}
\ee
and 
\be
\tilde n_i+\tilde n_j+\tilde n_k=0. \label{1.8}
\ee

There are various proposals to construct concretely these color-kinematic symmetric
numerators\footnote{See for example \cite{Mafra:2011kj,Broedel:2011pd,Fu:2012uy,Naculich:2014rta}.}.  
 However, in our view they are not straightforwardly implementable
via a set of conventional Feynman rules\footnote{ The exception is \cite{TW} who set out to derive BCJ numerators using a covariant (Feynman) gauge. In doing so they extended the particular effective five-point Lagrangian obtained by Bern et al in \cite{GGsq}. However their approach is somewhat less transparent than the steps we undertake in this paper and we were unable to see a direct translation of their algorithm into ours. }.  We will demonstrate that in fact
a general approach can be so prescribed.  We have chosen in what
follows to work in a light-like gauge of which space-cone gauge \cite{CS} is an example.  The reason is our
hidden desire to ultimately understand the connection of the gauge Lagrangian
to the gravity Lagrangian in some way.  The light-like gauges
seem to be the most promising, because explicitly there are
only two independent fields in each theory in four spacetime dimensions.  
Indeed, in \cite{AT-KLT}, by working in a light-like gauge, the authors were able to expose the squaring relation between the gravity and gauge theory four-point tree-level amplitudes, at the level of the Lagrangian. 
However, as we will
see, our procedure does not rely on any particular gauge choice, in the sense of  a specific choice of the light-like vector which can dramatically reduce the number of Feynman diagrams as  in \cite{CS}\footnote{Another benefit of the space-cone gauge is that it allows a straightforward proof of the BCFW on-shell recursion relations \cite{BCFW},  at the level of Feynman diagrams \cite{VY1}. By choosing the null space-cone gauge fixing vector such that it is expressed in terms of the two external gluon momenta which are analytically continued in the BCFW recursion, the only $z$-dependence in the analytically continued Feynman diagrams comes from the propagators. Then BCFW factorization is simply a statement about partial fractioning of the propagators in the Feynman diagrams followed by a regrouping into products of lower n-point amplitudes. In another application, the MHV Lagrangian was shown by Mansfield \cite{Mansfield} to be derived from a unitary transformation acting on the fields of the light-cone gauge fixed Lagrangian. Light-like gauges are useful beyond tree-level as well. We recall that Mandelstam used light-cone gauge for his proof of the UV finiteness of maximally supersymmetric Yang-Mills theories in four space-time dimensions \cite{Mandelstam}. On-shell recursion at one loop is also somewhat subtle, but space-cone gauge makes it for an easier approach \cite{VY2}.},  
if all we care is to obtain color-kinematic symmetric numerators. 

We have described how to relate the Kleiss-Kuijf set of amplitudes to
numerators which are color-kinematic symmetric.  When the numerators 
are not initially color-kinematic symmetric, as it is generally the case if we 
just apply Feynman rules as we normally would to calculate amplitudes,
then we must give a recipe how to modify them to make them
so.  The important criterion to observe is that the color-ordered amplitudes 
should be the same under such modifications.  To be more specific,
if we start out with
\be
n_i+n_j+n_k=\Delta _{ijk} \neq 0,  \label{1.9} 
\ee
we shall make changes
\be
n_l\to \bar n_l=n_l+\delta n_l \label{1.10}
\ee
such that
\be
\bar n_i+\bar n_j+\bar n_k=0,  \label{1.11}
\ee
which is equivalent to having the changes to absorb the
violation
\be
\delta  n_i+\delta  n_j+\delta  n_k=-\Delta _{ijk}.  \label{1.12} 
\ee
Now, we demand that from (\ref{1.1})
\be
\sum_i {c_i\delta n_i\over (\Pi s_j)_i}=0. \label{1.13}
\ee
It is easy to see that there are only $(n-2)!$ independent $\delta n_l$ and
upon expressing the others in terms of them and $\Delta$'s,  
we find that we end up with an equation
\be
|D\rangle =M|\delta N\rangle , \label{1.14}
\ee
in which $|\delta N\rangle $ is a column vector with the independent 
$\delta n$'s as entries, $|D\rangle $ is made of the $\Delta$'s, and 
$M$ is the same matrix as in (\ref{1.4}).

Just as before, because of the existence of eigenvectors with
null eigenvalue in $M$, we cannot invert the equation for $\delta n_l$
uniquely;
there are only $(n-3)!$ linear combinations of them which are active.
We must make some ansatz for the functional forms of these
$\delta n$'s and solve for them, which also points to the fact 
that there is in principle a whole host of choices one can make to render
the numerators dual symmetric.  What we would like to reiterate
is that the $\Delta$'s are constructed through Feynman rules.  They are uniquely given, once a gauge is picked.  On the other hand, there are
$(n-3)(n-3)!$ degrees of freedom in choosing $\delta n's$ ( and 
hence dual symmetric $n's$), which agree with the number
of generalized gauge transformations one can make.  By the same
token, there are  $(n-2)!$  entries in $D$, and we can use
any $(n-3)!$ of them for the 'inversion' of (\ref{1.14}). 

We will fix, in part, this freedom by requiring that the numerator shifts $\delta n$ do not introduce spurious poles. In other words, we require that the original pole structure expressed in writing the color-ordered amplitudes as in (\ref{1.1}), where $n_i$ are obtained via Feynman rules, is preserved. This will result in a tighter set of constraints imposed on the numerator shifts. For the five-points, the freedom in the numerator shifts reduces then to two arbitrary constants. As a consequence, we obtain color-symmetric numerators $\bar n_i$ in (\ref{1.4}) which will also preserve the original pole structure.

In the next few sections we will explicitly follow the program just outlined
for $n=4,\ 5$, to obtain a set of shifts which render the numerators 
color-kinematic symmetric.  We will write an effective Lagrangian for them.  The
parametrization of the other dual symmetric shifts will be given.
It will become obvious that the same procedure should work 
for any number of particles and in any light-like gauge.

The plan of this article is as follows. In Section \ref{section2} we give a quick overview of Yang-Mills theories in non-covariant, light-like gauges. We also introduce here our notation. In Section \ref{section3} we discuss four-point amplitudes as derived from the gauge-fixed Lagrangian. We notice that similar to results derived in covariant gauges, the numerators are already BCJ symmetric. The next two sections, \ref{section4} and \ref{section5}, are dedicated to the five-point amplitudes and the corresponding numerators. The numerators obtained via Feynman rules are not BCJ symmetric. However, we show that there is an effective null five-point Lagrangian which induces shifts of the numerators such that the end result is BCJ symmetric. We relegate technical details and intermediate results to four of the appendices.  We make some final remarks in Section \ref{section6} and comment on extending our procedure to six-point functions in Appendix \ref{Appendix5}.

\section{Notation, conventions, and a quick overview of light-like
gauges:}\label{section2}
Throughout this paper we work in four space-time dimensions.
The Lorentz metric we use is defined via the scalar product
\be
P_\mu Q^\mu=-P^0 Q^0+\vec P \cdot \vec Q
= p\bar q +\bar p q-p^+q^- -p^-q^+, \label{2.1}
\ee
where we have introduced the notation
\be
p^\pm\equiv \frac{1}{\sqrt 2}(P^0\pm P^3), \qquad p\equiv \frac1{\sqrt 2}
(P^1+i P^2), \qquad \bar p\equiv\frac 1{\sqrt 2}(P^1-i P^2).\label{2.1a}
\ee
We reserve capital letter notation for
vectors carrying Greek indices: $P_\mu=(P_0, \vec P)$.

Following \cite{CS}, we introduce the reference (commuting) spinors $|\pm\rangle$ and
$|\pm]$, normalized to
\be
\langle+-\rangle=[-+]=1  \label{2.2}
\ee
but otherwise arbitrary.
Then the set of null vectors $\{|+\rangle [ + |, |-\rangle [+|, 
|+\rangle [-|, |-\rangle [-|\}$ forms a basis and the four-vector components introduced earlier in  (\ref{2.1a}) are obtained from the 
decomposition
\be
P=p^+ |+\rangle [ + |\,+\, p^- |-\rangle [-|\,+\, 
p |+\rangle [-|\,+\,\bar p |-\rangle [+|.
\ee
If $P^\mu$ is a null four-vector, i.e. there exist spinors such that 
$P=|p\rangle [p|$, then
\be
p=\langle p+\rangle [p-], \ \bar  p=\langle p-\rangle [p+], \ p^+=\langle p-\rangle [-p], \ 
p^-=\langle p+\rangle[+p]. \label{2.3}
\ee

Starting with the Yang-Mills Lagrangian
\be
{\cal L}=-\frac 14 F_{\mu\nu \,a}{F^{\mu\nu}}_a
\ee
where $a$ is an adjoint color index and 
the field strength $F_{\mu\nu\; a}$ is given as
\be
F_{\mu\nu\,a}=\partial_{\mu} A_{\nu\,a}-\partial_\nu A_{\mu\,a} +g f_{abc} A_{\mu \,b} A_{\nu\, c},
\ee
one can reach the non-covariant gauge \cite{CS}
\be
a_b=0\label{2.5}.
\ee
This is analogous to the more familiar light-cone gauge fixing condition $a^+_b=0$. 
Both gauges are light-like, in the sense that one sets to zero a component of the gauge field along a given null vector.

In components, the  Yang-Mills Lagrangian is
\bea
{\cal L}&=&-\frac 14\bigg[2 (\partial^+ a^{-}_{a} - \partial^- a^{+}_{a}+
g f_{abc} a^{+}_{b} a^{-}_{c})(\partial^- a^{+}_{a} - \partial^+ a^{-}_{a}+
g f_{ade} a^{-}_{d} a^{+}_{e})\nonumber\\
&&-4(\partial^+ a_{a}-\partial a^{+}_{a} + g f_{abc} a^{+}_{b} a_c)(
\partial^- \bar a_{a}-\bar \partial a^{-}_{a} + g f_{ade} a^{-}_{d} \bar a_e)
\nonumber\\
&&-4(\partial^+ \bar a_{a}-\bar \partial a^{+}_{a} + g f_{abc} a^{+}_{b} 
\bar a_c)(\partial^- a_{a}- \partial a^{-}_{a} + g f_{ade} a^{-}_{d}  a_e)\nonumber\\
&&+2(\partial \bar a_{a}-\bar\partial a_{a}+g f_{abc} a_b\bar a_c)( \bar\partial a_{a}-\partial \bar a_{a}+g f_{ade} \bar a_d a_e)\bigg]
\eea 
where all derivatives are understood to be $\partial^\mu$.
For example, $\partial^+=-\frac{\partial}{\partial x^-}$, $\partial=\frac{\partial}{\partial \bar x}$ etc. In momentum space these derivatives convert simply to factors of the corresponding momentum components: $\partial^+$ becomes $ i p^+$, $\partial$ becomes $ip$ etc.

After using the gauge fixing condition $a_b=0$, $\bar a_b$ is independent of the ``time''-derivative $\bar\partial$ and so it can be eliminated from its equation of motion, 
\be
\bar a_b=\frac1{\partial}\bigg[\partial^+ a^{-}_{b} + \partial^- a^{+}_{ b} - g \frac{f_{bcd}}{\partial}(\partial a^{-}_{c}a^{+}_{d}+\partial a^{+}_{c}a^{-}_{d})\bigg]
\ee
The gauge fixed Lagrangian
becomes
\bea {\cal L} &=&-a_a^-\partial_\mu \partial^\mu a_a^+
+ 2 g f_{abc}({\partial^+\over \partial}a_a^-)a_b^-\partial a_c^+
+ 2 g f_{abc}({\partial^-\over \partial}a_a^+)a_b^+\partial a_c^-\nonumber \\
&&
+ 2 g^2 (f_{abc}a_b^-\partial a_c^+){1\over \partial ^2}
  (f_{ade}a_d^+\partial a_e^-). \label{2.6}
\eea
This Lagrangian contains now only the two physical degrees of freedom of a gauge field in four space-time dimensions: positive and negative helicities corresponding respectively to the $a^+$ and $a^-$ components.\footnote{Note that in \cite{KS}, the gauge-fixed Lagrangian given in (14) has a sign typo  in  the kinetic term. For another comparison, \cite{CS} have their Lagrangian being normalized as ${\cal L}=\tfrac 1{8g^2} Tr(F_{\mu\nu} F^{\mu\nu})$ with $F_{\mu\nu}\equiv F_{\mu\nu\,a} T_a = (\partial_\mu A_{\nu\,a} -\partial_\nu A_{\mu\,a}) T_a + A_{\mu\,b} A_{\nu\,c}\,[T_b, T_c]$. After rescaling the gauge fields $A_{\mu \,a}\to g A_{\mu\,a}$, and using that in the adjoint representation the gauge group generators equal $(T_a)_{bc}=i f_{abc}$ where $f_{abc}$ are the structure constants ($[T_a, T_b]=-if_{abc} T_c$), their Lagrangian becomes our (\ref{2.6}) up to an overall factor $-\tfrac 12$ times a normalization factor $N$ obtained from the evaluation of the traces  $Tr(T_a T_b)=N\delta_{ab}$. Taking $N=2$ results in agreement between the Lagrangian in \cite{CS} and (\ref{2.6}) up to an overall sign.}

This yields the following Feynman rules:

\begin{figure}[h]
Propagator:   $\qquad$ \includegraphics[scale=0.25]{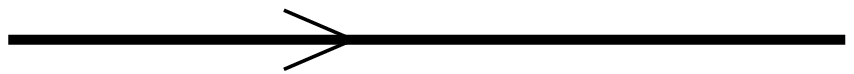}
$\;=\;\qquad\frac{-i\delta_{ab}}{P^2}$\end {figure}

\begin{figure}[h]
Three-point  vertices:
$\;\;\;$
\includegraphics[scale=0.25]{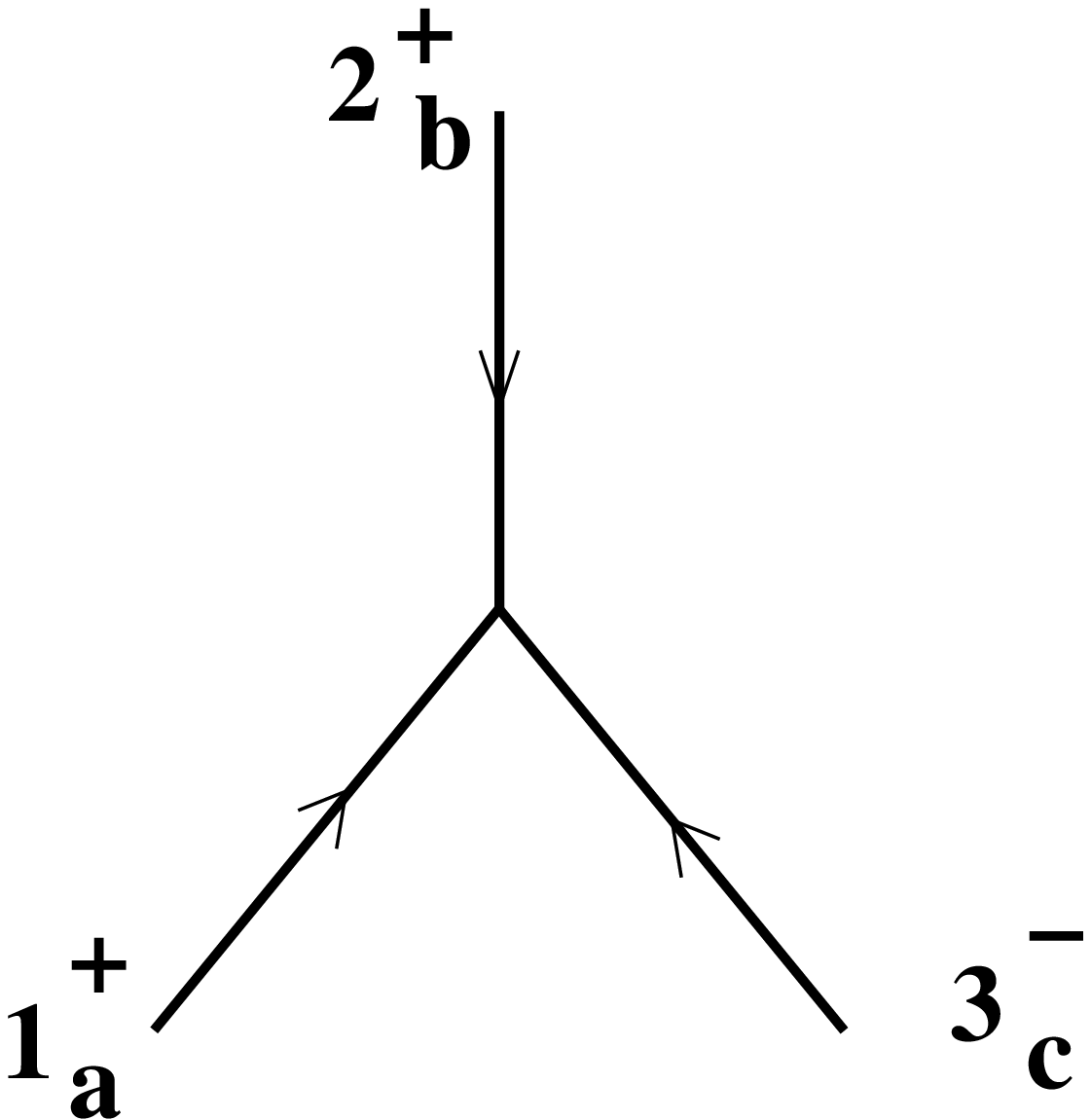}
$\;=\;- 2gf_{abc} ({p_1^-\over p_1}-{p_2^-\over p_2})p_3$\\
\\
 
$\;\;\;\;\;\;\;\;\;\;\qquad\qquad\qquad\;\;$\includegraphics[scale=0.25]{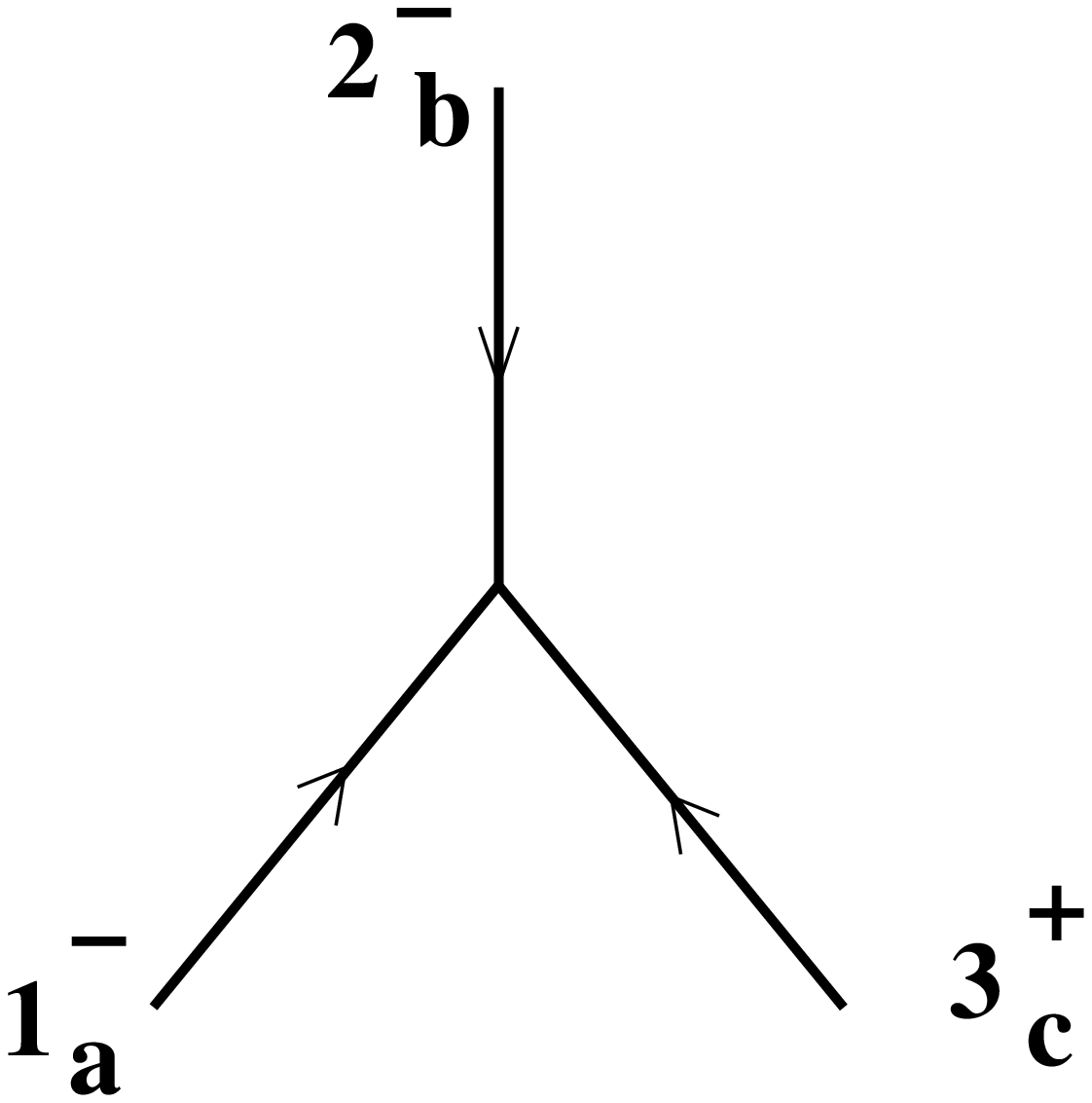}
 $\;=\;-2g f_{abc}({p_1^+\over p_1}-{p_2^+\over p_2})p_3$
\end{figure}

\newpage Four-point vertices\footnote{We choose to interpret the quartic term in the Lagrangian
$ 2 g^2 (f_{abc}a_b^-\partial a_c^+){1\over \partial ^2}
  (f_{ade}a_d^+\partial a_e^-)$ as $-
2 g^2 \partial_{\mu}(f_{abc}a_b^-\partial a_c^+){1\over \Box\partial ^2}
 \partial^\mu (f_{ade}a_d^+\partial a_e^-) $. The manifest propagator in the denominator makes it clear how we choose to assign the contribution of the four-point vertex to the numerators. We note that a similar choice was made by \cite{TW}.}:

\begin{figure}[h]
\includegraphics[scale=0.25]{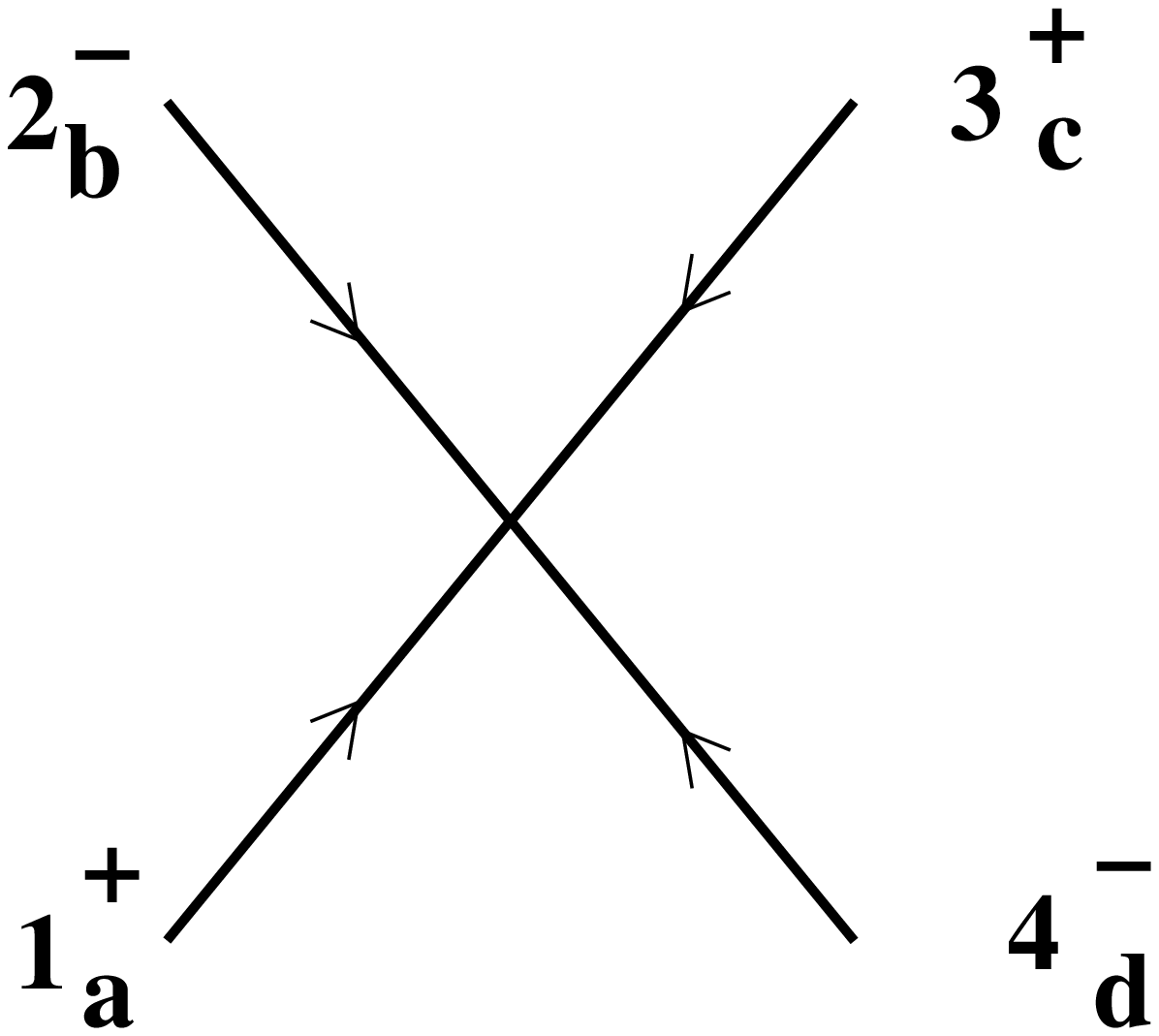}
\includegraphics[scale=0.25]{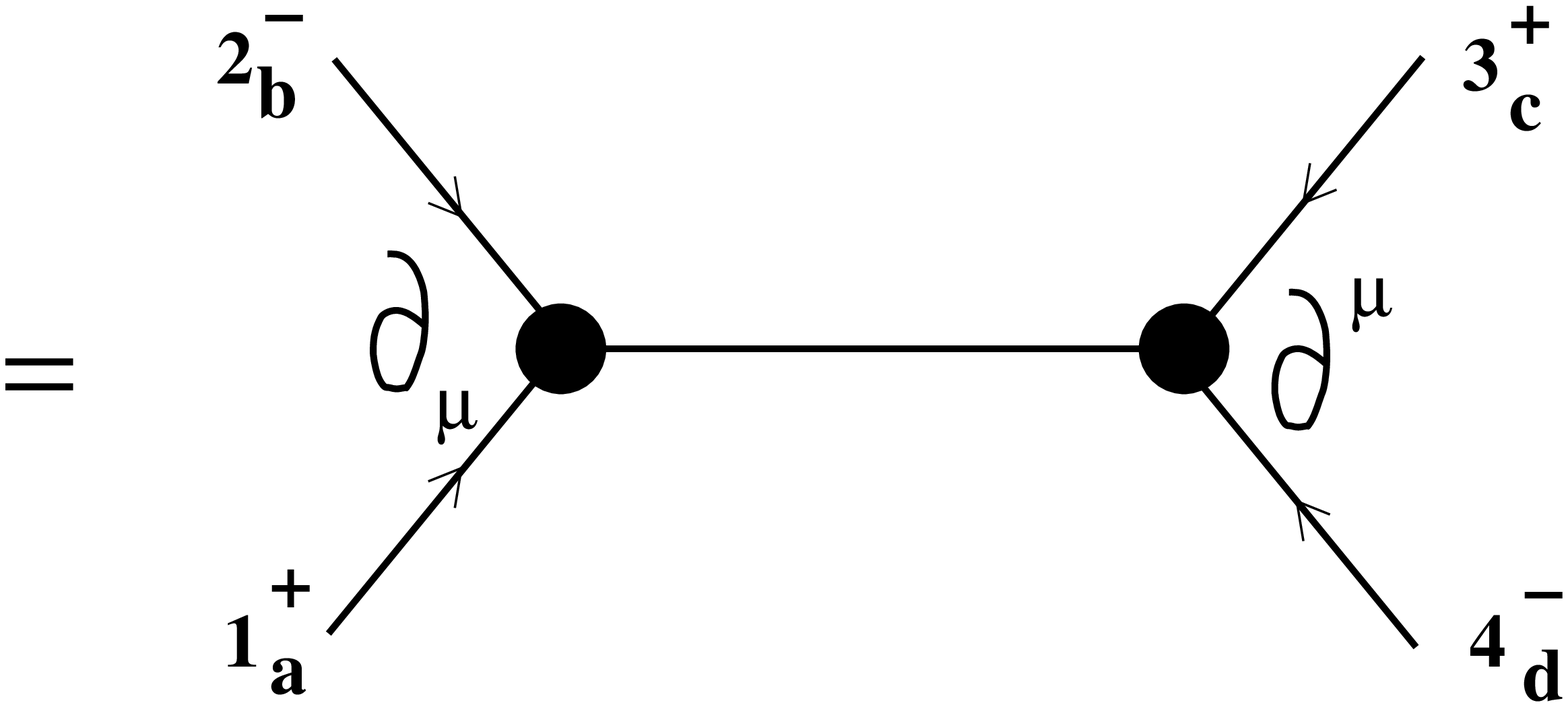}
\includegraphics[scale=0.25]{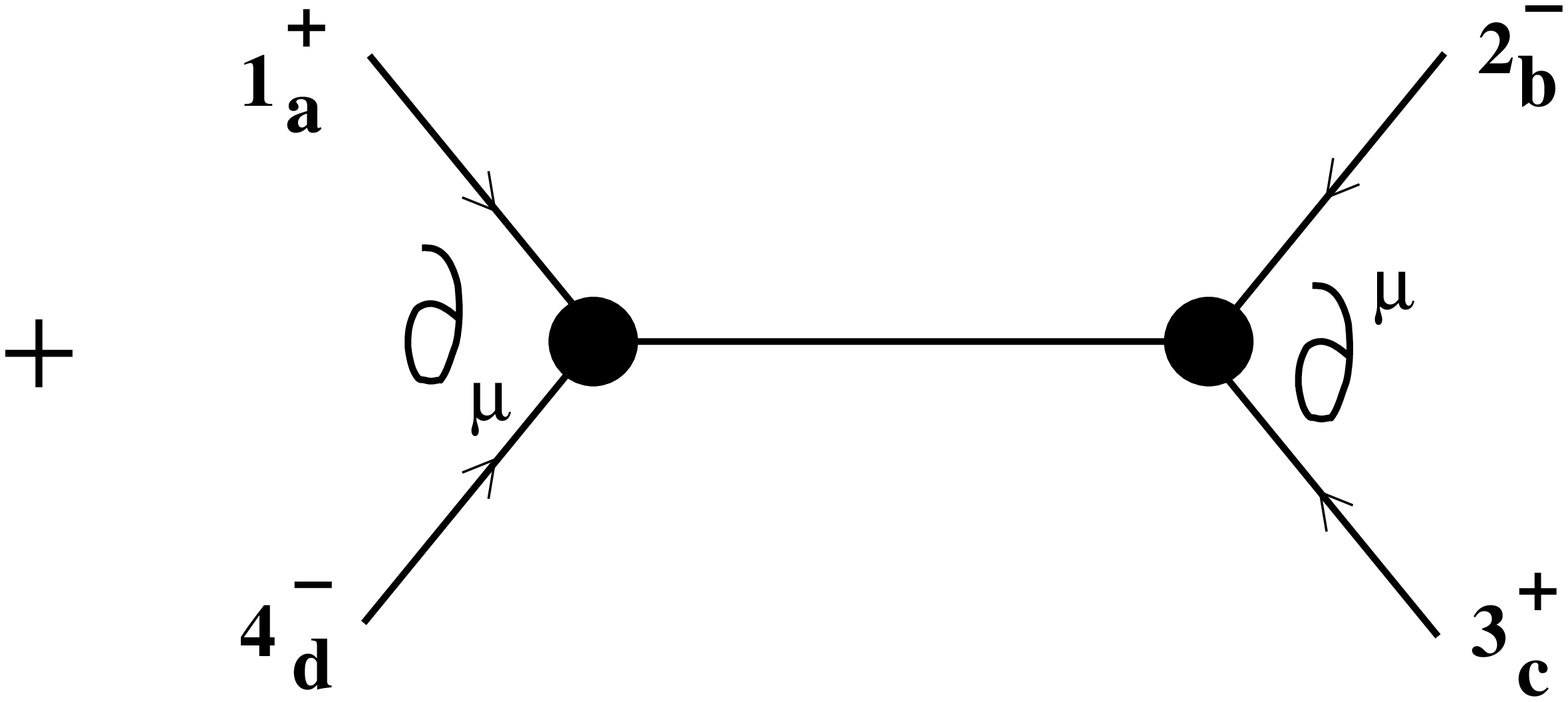}\\

$\;\;\;\;\;\;\;\;\;\;\;\;\;\;\;\;\;\;\;\;\;\;\;\;\;\;\;=\qquad\qquad -2g^2 i f_{abe}f_{cde}\frac{p_1p_4+p_2p_3}{(p_1+p_2)^2}\frac{s_{12}}{s_{12}} \;+\; (-2g^2 i)  f_{dae}f_{bce} \frac{p_1p_2+p_3p_4}{(p_1+p_4)^2}\frac{s_{14}}{s_{14}}. $


\end{figure}

 These Feynman rules need to be supplemented by the insertion of external line factors. These originate in the polarization vectors and their components corresponding to the positive or negative helicity of a given external line. More concretely, 
\be
\epsilon(P) ^+={[-p]\over \langle +p\rangle}, \ 
\epsilon(P)^-={\langle +p\rangle \over [-p]},  \label{2.4}
\ee
need to be inserted for each external line with momentum $P$ and positive or negative helicity respectively.

Chalmers and Siegel noted the advantage which comes from choosing the reference spinors in such a way that $|+\rangle[+|$ is the momentum of an external negative helicity gluon and $|-\rangle [-|$ is the momentum of an external positive helicity gluon (up to a normalization factor). This choice leads to the least number of Feynman diagrams for a given process (far less then in other more covariant gauges), and they referred to it as space-cone gauge \cite{CS}. 
 However, here we work in full generality, and keep the reference spinors arbitrary. To emphasize this distinction we will refer to the gauge condition in (\ref{2.5}) as a light-like gauge condition.

It should be pointed out that in a light-like gauge, where
we separate out particles of two different helicities,
the symmetries among like helicity particles are explicit, while
those between unlikes must be imposed by hand.  

 In the usual fashion, we will convert the structure constant factors into traces over the group generators, and compute color ordered amplitudes.
 The tree level $n$-gluon scattering amplitude is then equal to the sum over the color-ordered partial amplitudes 
\be
A^{(n)}=-i g^{n-2} \sum_{\sigma} Tr[T_{a(\sigma(1))} T_{a(\sigma(2))} \dots T_{a(\sigma(n-1))}
T_{a(\sigma(n))}] A(\sigma(1),\sigma(2),\dots\sigma( n)),\label{tree color}
\ee  where $\sigma$ is a non-cyclic permutation of the external gluons. 

To identify a given numerator by the labelling of the indices, we 
follow the convention we developed in \cite{VY3}.  Thus, for
$n$ gluons, we have a set of color ordered numerators.
As a consequence of clockwise vs. counterclockwise tracing, the numerators 
satisfy
\be 
n(i_n \cdots i_1)=(-1)^nn(i_1\cdots i_n),\label{2.9}
\ee

The indices can be further refined: if two adjacent indices,
say $j$ and $j+1$, share the same structure constant 
$f_{a_ja_ {j+1}a_k}$, we shall separate them from the
other indices by two sets of semi-colons 
$n(i_1\cdots i_{j-1};i_j i_{j+1};i_{j+2}\cdots i_n)$.  Clearly they 
are antisymmetric
\be
n(i_1\cdots i_{j-1};i_j i_{j+1};i_{j+2}\cdots i_n)=
-n(i_1\cdots i_{j-1};i_{j +1}i_j;i_{j+2}\cdots i_n), \label{2.10}
\ee
because of the form taken by the cubic and quartic vertices.

It is also clear that for each pair of such indices, 
a propagator ${i\over 2s_{j \ j+1} }$ will go with them
in an amplitude, where $s_{j \ j+1}=-\tfrac 12(P_j+P_{j+1})^2$.  For example in writing $n(i_1 i_2;i_3;\dots;i_{n-1}i_n)$, the pole structure associated with it is $s_{i_1 i_2}=-\tfrac 12(P_{i_1}+P_{i_2})^2$, $s_{i_1 i_2 i_3}=-\tfrac 12(P_{i_1}+P_{i_2}+P_{i_3})^2,\dots$ $s_{i_1 i_2 i_3 \dots i_{n-2}}=-\tfrac 12(P_{i_{n}}+P_{i_{n-1}})^2$.

\section{ Duality for Four Particles:}\label{section3}

In this section we carry out the program outlined in the Introduction for the simplest
case when  $n=4$.  As it is well-known, the configurations
$\pm\pm\pm\pm$ and $\pm\pm\pm\mp $
are trivial, because the amplitudes vanish.  The Jacobi permutation 
of the numerators $n_i+n_j+n_k$
 vanishes even off-shell.  For the maximal helicity violation
case $1^+2^-3^+4^-$ and other helicity assignments, we 
show that when the particles are all on shell the numerators are
automatically dual symmetric, if we just apply the Feynman rules
to obtain them in any light-like gauge, 
and particularly in the space-cone gauge.
We then extend this to obtain a result for the relevant Jacobi
cyclic permutation  when the particles are off-shell,
which will be used in the next section as an insertion.  We find  here
that each term is proportional to the invariant mass of one of the
four particles, which is an affirmation that the numerators are BCJ symmetric on-shell, as said already mentioned.

Upon using (\ref{2.9}) and (\ref{2.10}), we see that it is sufficient
to deal with the numerators $n(12;34), \ n(13;24),$ and $n(23;14)$.
Let us focus on cyclically permuting the first three indices, 
\be
n(12;34)+n(23;14)+n(31;24)\equiv \Delta(123|4), \label{3.1}
\ee
or
\be
n(23;41)=n(12;34)-n(13;24)-\Delta(123|4). \label{3.2}
\ee
The last equation means that $n(12;34)$ and $n(13;24)$ can
be taken as the independent numerators, while $n(23;41)$ being 
given by them and the amount of duality violation $\Delta (123|4)$.
Similarly, if we Jacobi permute the last three indices
\be
n(12;34)+n(13;42)+n(14;23)\equiv \Delta (1|234), \label{3.3}
\ee
we see that using (\ref{2.9}) and (\ref{2.10}), we are yielded
\be
\Delta (123|4)=\Delta (1|234), \label{3.4}
\ee

For the color-ordered amplitudes in the Kleiss-Kuijf basis, chosen for concreteness to be composed of $A(1234)$ and $A(1324)$, we have
\bea
A(1234)  &=&{n(12;34)\over s_{12}}+{n(23;41)\over s_{14}}\nonumber\\
&=&n(12;34)({1\over s_{12}}+{1\over s_{14}})+n(13;24)(-{1\over s_{14}})
+\Delta (123|4)(-{1\over s_{14}}), \label{3.5}
\eea
and
\bea
A(1324)&=&{n(13;24)\over s_{13}}-{n(23;41)\over s_{14}}\nonumber\\ 
&=&n(12;34)(-{1\over s_{14}})+n(13;24)({1\over s_{13}}+{1\over s_{14}})
+\Delta (123|4)({1\over s_{14}}). \label{3.6}
\eea

The next step is to modify the numerators derived from the use of Feynman rules by adding $\delta n$ terms such that the resulting $\bar n= n + \delta n$ numerators obey Jacobi identity, and such that the amplitudes are unchanged.
More concretely,
\bea
&&\bar n(12;34)=n(12;34)+ \delta n(12;34), \ 
\bar n(13;42)=n(13;42)+\delta n(13;42),\nonumber\\ &&
\bar n(14;23)=n(14;23)+\delta n(14;23),  \label{3.7}
\eea
are defined so that 
\be
\bar n(12;34)+\bar n(13;42)+\bar n(14;23)=0, \label{3.8}
\ee
or
\be
\delta n(12;34)+\delta n(13;42)+\delta n(14;23)=
-\Delta (123/4), \label{3.9}
\ee
such that the values of the color-ordered amplitudes are not changed.  
Please note that by definition we are referring to on-shell quantities
here.  When we extend the amplitudes to amputated Green's 
functions, we cannot make such a demand.
From the requirement that the change made to the numerators does not change the amplitudes, which in terms of the color-kinematic symmetric numerators $\bar n$ are written as  
\bea
&&\begin{pmatrix} A(1234) \\ A(1324) \end{pmatrix} 
= M^{(4)}
\begin{pmatrix}  \bar n(12;34) \\ \bar n(13;24) \end{pmatrix},
 \eea
we are led to the following constraint on the numerator shifts in the chosen Kleiss-Kuijf basis:
\bea 
\begin{pmatrix} {-\Delta(123|4) \over s_{14}} \\ {\Delta(123|4) \over s_{14}} \end{pmatrix} 
= M^{(4)}
\begin{pmatrix} \delta n(12;34) \\ \delta n(13;24) \end{pmatrix}, \label{3.10}
\eea
with $M^{(4)}$ the four-point propagator matrix introduced in \cite{VY3}
\bea
M^{(4)}=\begin{pmatrix}  {1\over s_{12}}+{1\over s_{14}}   &
 -{1\over s_{14}} \\
-{1\over s_{14}} & {1\over s_{13}}+ {1\over s_{14}} \end{pmatrix}.\label{3.11}
\eea
An important observation made in \cite{VY3} is  that $M^{(4)}$ has an eigenvector with
zero eigenvalue
\be
\langle \lambda^0|=\langle -s_{12},\ s_{13}|.  \label{3.12}
\ee
Then one has the freedom to change the numerators by adding these zero eigenvectors. In doing so, the defining equation (\ref{3.11}) remains the same. This freedom was called generalized gauge transformation in \cite{VY3}.\footnote{We would like to emphasize that the shifts $\delta n$ cannot be obtained in general by making generalized gauge transformations. The four-point case is somewhat special since we will argue that the numerators satisfy the color-kinematic duality without any need to make these shifts. However, this does not extend to the higher n-point numerators. } For the four point amplitudes, the implication is that there is only one effective $\bar n$ and
one effective $\delta n$.  For the latter, we make the following generalized gauge transformation
\be
\begin{pmatrix} \delta n(12;34) \\ \delta n(13;24) \end{pmatrix}
\to \begin{pmatrix} \delta n(12;34) \\  \delta n(13;24) \end{pmatrix}
-{\delta n(13;24)\over s_{13}}|\lambda ^0\rangle =\begin{pmatrix} \delta n \\ 
0 \end{pmatrix} \label{3.13}
\ee
where
\be
\delta n\equiv\delta n(12;34)+{s_{12}\over s_{13}}\delta n(13;24).
\label{3.14}
\ee
This results in a reduced equation
\be
\begin{pmatrix} {-\Delta(123|4) \over s_{14}}\\{\Delta(123|4) \over s_{14}} \end{pmatrix} 
= M
\begin{pmatrix} \delta n \\ 0 \end{pmatrix}, \label{3.15}
\ee
which demands
\be
\Delta (123|4)=0.  \label{3.16}
\ee
We will verify this explicitly in a direct calculation below.  In the mean
time, it tells us that the numerators calculated through Feynman
rules for the on shell $n=4$ amplitudes are
dual symmetric without any need for modification.  We should 
point out that up to this point, we need not refer to any specific choice of gauge, light-like or covariant,  
to come to this conclusion.

Let us turn to the off-shell situation, by which we mean of course
that the invariant mass of each individual particle is non-zero. Also,
we do not let the numerator matrix elements act on the polarization tensors.  Since we have
\be
n(12;34)+n(31;24)+n(23;14)=n(12;34)+n(14;23)+n(13;42) \label{3.17}
\ee
there is only one Jacobi permutation.  Also, for a given number of 
$+$ and $-$, the specific assignment to each individual 
particle can be arbitrary, because we cycle them through the
permutations above, which will cover all the cases if we relabel the
particle number.  For
$\pm \pm \pm \pm$, the case is trivial, because the vertices cannot
be matched to make the scattering go.  For $\pm \pm \pm \mp$,
we need only three-point vertices of the same type $(++-)$. In the case we are considering the numerators are
\bea
&& n(1^+2^+;3^+4^-)=({p_1^-\over p_1}-{p_2^-\over p_2})(-(p_1+p_2))
({p_1^-+p_2^-\over p_1+p_2}-{p_3^-\over p_3})p_4 , \nonumber\\&&
n(1^+4^-;2^+3^+)=
({p_2^-\over p_2}-{p_3^-\over p_3})(-(p_2+p_3))({p_2^-+p_3^-\over p_2+p_3}-{p_1^-\over p_1})p_4 , \nonumber\\ &&
n(1^+3^+;4^-2^+)=({p_3^-\over p_3}-{p_1^-\over p_1})(-(p_3+p_1))
({p_3^-+p_1^-\over p_3+p_1}-{p_2^-\over p_2})p_4 . \label{3.18}\eea
When we add them, we find that all terms in the sum cancel
completely.

The $\pm \pm \mp \mp$ case is the non-trivial one.
For one thing, four-vertices make their appearance.  In view of the
somewhat tedious algebra to bring the expressions to the final form, we are relegating the details to Appendix \ref{Appendix1}.  The results are
\bea
n(1^-2^+;3^+4^-)&=&s_{23}+({p_2^-\over p_2}-{p_3^-\over p_3})
({p_4^+\over p_4}-{p_1^+\over p_1})(p_1 p_3+p_2 p_4)\nonumber\\ &&
+{p_1p_3\over p_1+p_2}(-{P_4^2\over 2 p_4}+{P_2^2\over 2 p_2})
+{p_2p_4\over p_1+p_2}(-{P_3^2\over 2 p_3}+{P_1^2\over 2 p_1}),
 \label{3.19}\\
n(3^+1^-;2^+4^-)&=&-s_{23}+({p_2^-\over p_2}-{p_3^-\over p_3})
({p_4^+\over p_4}-{p_1^+\over p_1})(p_1 p_2+p_3 p_4)\nonumber\\ &&
+{p_3p_4\over p_2+p_4}(-{ P_2^2\over 2 p_2}+{P_1^2\over 2 p_1})
+{p_1p_2\over p_2+p_4}(-{ P_4^2\over 2 p_4}+{P_3^2\over 2 p_3}),
 \label{3.20}
\eea
and 
\be
n(2^+3^+;1^-4^-)=-({p_2^-\over p_2}-{p_3^-\over p_3})
({p_4^+\over p_4}-{p_1^+\over p_1})(p_1 +p_4)(p_2+ p_3),
 \label{3.21}
\ee
where we have omitted a product of the four polarization vectors
because we are extending the result to off-shell $P_i^2=\vec P_i^2-(P_i^0)^2
\ne 0 $.  We now add them and find
\bea
\Delta (2^+|1^-3^+4^-)&\equiv& n(2^+1^-;3^+4^-)+n(2^+3^+;4^-1^-)
+n(2^+4^-;1^-3^+)\nonumber\\ &
=&  \tfrac 12 P_1^2p_4({1\over p_1+p_2}-{1\over p_1+p_3})
+  \tfrac 12 P_2^2p_3({1\over p_1+p_2}-{1\over p_2+p_4})\nonumber\\ &&  
+ \tfrac 12 P_3^2p_2({1\over p_3+p_4}-{1\over p_1+p_3})
+ \tfrac 12 P_4^2p_1({1\over p_3+p_4}-{1\over p_2+p_4}).\label{3.22}
\eea 
When we go on-shell, by setting $P_i^2
\to 0$, we have $\Delta \to 0$, which, as advertised, means that
duality holds by the on-shell numerators as calculated through regular Feynman rules, without any need for additional adjustment.  We will 
find the off-shell $\Delta (2^+|1^-3^+4^-)$ useful as an insertion in the next section
when we look into the five particle case.  
The fact that it is non-vanishing is an indication that it is
non-trivial in 
enforcing dual symmetry for higher point numerators.  
As a reminder, the off-shell parts of (\ref{3.19}-\ref{3.20})$\approx P_i^2$ can be associated with an operator insertion 
\bea
&&f_{bac}f_{b'a'c}[({1\over \Box }{1\over \partial}
(\partial a^-_b { \Box\over \partial}a^+_a))
(\partial a^+_{b'}a^-_{a'})
-({1\over \Box }{1\over \partial}
(\partial a^-_b a^+_a))
(\partial a^+_{b'}{ \Box\over \partial}a^-_{a'})]\nonumber\\
&&\propto Tr \bigg(\frac{1}{\partial}\frac{1}{\Box}([a^-,\partial a^+])[\partial a^-,\frac{\Box}{\partial}a^+]\bigg),
\label{3.23}
\eea
which generates the off-shell $\Delta (2^+|1^-3^+4^-)$ and others
with $\Box=\partial^\mu \partial_\mu$.\footnote{We note in passing that the operator insertion (\ref{3.23}) which insures that the off-shell four-point enjoys the color-kinematic duality can be obtained via the following field redefinition
$a^-_b \to a^-_b - \frac 1{\partial}\bigg(
f_{bcd}f_{deg}\partial a^-_c \frac{1}{\partial\Box}(a^-_e\partial a^+_g)\bigg),$ and its parity conjugate counterpart. }

For completeness, let us use these numerators to calculate (and check) one of
the color ordered amplitudes.  It helps to note that when on-shell
\be
s_{ij}=-p_ip_j({p_i^+\over p_i}-{p_j^+\over p_j})
({p_i^-\over p_i}-{p_j^-\over p_j}), \label{3.24}
\ee
and
\be
{p_a^+\over p_a}-{p_b^+\over p_b}={\langle ab\rangle \langle +-\rangle \over \langle +a\rangle \langle +b\rangle },
\ {p_a^-\over p_a}-{p_b^-\over p_b}={[ab][-+]\over [-a][-b]},
\label{3.25}
\ee
Then some simple algebra gives
\bea
A(1^-2^+3^+4^-)&=&{n(1^-2^+;3^+4^-)\over s_{12}}-
{n(2^+3^+,1^-4^-)\over s_{14}}\nonumber\\ &
=&-{p_1p_2p_3p_4\over s_{12}s_{14}}({p_2^-\over p_2}-{p_3^-\over p_3})^2
({p_4^+\over p_4}-{p_1^+\over p_1})^2\epsilon _1^-
\epsilon _2^+ \epsilon _3^+ \epsilon_4^- \nonumber\\ &
=&{\langle14\rangle^4\over \langle12\rangle\langle23\rangle\langle34\rangle\langle41\rangle} \label{3.26}
\eea
a well-known result.

We did not make any choice of the reference vectors $|+\rangle[+|$ or $|-\rangle[-|$
up to this point in order to show generality.  However, if the intention is to
shorten a calculation, then some particular choices can be expeditious.
For example, if we take $|1\rangle[1|\propto|+\rangle[+|$ and $|2\rangle[2|\propto |-\rangle[-|$, we have
$\epsilon _1^-, \ \epsilon_2^+, \ p_1, \ p_2 \to 0$.  However
\be
\epsilon_1^-{p_1^+\over p_1} \to 1, \  \epsilon_2^+{p_2^-\over p_2}
\to 1, \label{3.27}
\ee
and many terms can be dropped to give immediately
\be
A(1^-2^+3^+4^-)=-\epsilon_3^+\epsilon_4^-{p_3p_4\over s_{14}}
={\langle14\rangle^4\over \langle12\rangle\langle23\rangle\langle34\rangle\langle41\rangle} .\label{3.28}
\ee

\section{ Duality for Five Particles:}\label{section4}

For the five particle amplitudes following \cite{VY3} we choose the Kleiss-Kuijf basis to be composed of $A(12345), A(14325), A(13425), A(12435), A(14235), A(13245)$. Each of these amplitudes has simple poles in the various kinematic invariants. There are fifteen numerators associated with these poles, owing to 
symmetries such as (\ref{2.9}) and (\ref{2.10}). 
 Keeping the same notation as in \cite{VY3} we  denote six of them as follows\footnote{If we use color-kinematics duality, these six numerators would be the independent set in terms of which all others are expressed. However, here we are concerned with a Lagrangian-based approach, and as we will see the numerators obtained via Feynman diagrams in a generic light-like gauge do not obey color-kinematics duality. We denote the violation of color-kinematic duality by $\Delta$ and we compute the specific $\Delta$'s. Only after modifying the numerators by $\delta n$ shifts will the resulting numerators obey color-kinematic duality. Of course the shifts are required to leave the amplitudes unchanged, as we did in the previous section.}:
\bea
&& n_{1}=n(12;3;45), \ n_{12}=n(12;4;35), \ n_{15}=n(13;2;45), \nonumber\\ &&
n_9=n(13;4;25), \ n_{14}=n(14;2;35), \ n_6=n(14;3;25). \label{4.1}
\eea

 Then we incorporate Jacobi permutations of the last three indices to express 
\bea
 && n(12;5;34)=-n_1+n_{12}+\Delta (12|345),\nonumber \\&&
n(13;5;24)=-n_{15}+n_9+\Delta (13|245),\nonumber \\&&
n(14;5;23)=-n_{14}+n_6+\Delta (14|352),\nonumber \\&&
n(15;2;34)=-n_1+n_{12}+n_9-n_6+\Delta (12|345)+\Delta (25|134)
+\Delta (34|125),\nonumber \\&&
n(15;3;42)=-n_{12}+n_{15}-n_9+n_{14}-\Delta (35|124)-\Delta (24|135)
-\Delta (13|245),\nonumber \\&&
n(15;4;23)=n_1-n_{15}-n_{14}+n_6+\Delta (14|352)+\Delta (45|123)
+\Delta (23|145),\nonumber \\&&
n(23;1;45)=-n_1+n_{15}-\Delta (45|123),\nonumber \\&&
n(24;1;35)=-n_{12}+n_{14}-\Delta (35|124),\nonumber\\&&
n(25;1;34)=n_9-n_6+\Delta (25|134).\label{4.2}
\eea
We will later give concrete expressions for the $\Delta $'s for the 
configuration $1^+2^-3^+4^-5^+$.  Actually there is one
extra equation which over-determines the quantities in (\ref{4.2}).
Thus, for consistency, one has to have
\bea
\Delta (13|245)&=&\Delta (45|123)+\Delta (23|145)
+\Delta (34|125)\nonumber\\
 &+&\Delta (12|345)
+\Delta (25|134)+\Delta (14|352) \nonumber\\ &
-&\Delta (35|124) -\Delta (24|135)-\Delta (15|234), \label{4.3}
\eea
which will be checked.

Then color-kinematic duality statement is that there is a set of numerators $\bar n$'s, obeying Jacobi identity under cyclic permutation of three indices.
The algebraic relation between the amplitudes and the BCJ numerators is
\be
\begin{pmatrix} A(12345)\\  A(14325) \\ A(13425) \\ 
A(12435) \\ A(14235) \\ A(13245) \end{pmatrix}=
M^{(5)}
\begin{pmatrix}  \bar n(12;3;45) \\ \bar n(14;3;25) \\ \bar n(13;4;25) \\ 
 \bar n(12;4;35) \\ \bar n(14;2;35) \\  \bar n(13;2;45) \end{pmatrix}
=
M^{(5)}
\begin{pmatrix}  \bar n_1 \\ \bar n_6 \\ \bar n_9 \\ 
 \bar n_{12} \\ \bar n_{14} \\  \bar n_{15} \end{pmatrix}, \label{4.9}
\ee
where the propagator matrix $M^{(5)}$ is given by the following:
\bea
&&
\!\!\!\!\!\!\!\!\!\!\begin{pmatrix}
\frac{1}{s_{12}s_{45}}+\frac{1}{s_{15}s_{34}}&\frac{1}{s_{15}s_{34}}+\frac{1}{s_{23}s_{15}}&-\frac{1}{s_{15}s_{34}}&-\frac{1}{s_{15}s_{34}}-\frac1{s_{12}s_{34}}&
-\frac{1}{s_{23}s_{15}}&-\frac{1}{s_{23}s_{45}}-\frac{1}{s_{23}s_{15}}\\
+\frac{1}{s_{23}s_{15}}+\frac{1}{s_{12}s_{34}}& & & & &\\
+\frac{1}{s_{23}s_{45}}& & & & & \\
\\
\frac{1}{s_{15}s_{34}} +\frac{1}{s_{15}s_{23}}&\frac{1}{s_{14}s_{25}}
+\frac{1}{s_{14}s_{23}} &-\frac{1}{s_{15}s_{34}} -\frac{1}{s_{34}s_{25}}&
-\frac{1}{s_{15}s_{34}} & -\frac{1}{s_{14}s_{23}}-\frac{1}{s_{15}s_{23}}&
-\frac{1}{s_{15}s_{23}}\\
&+\frac{1}{s_{15}s_{23}}+\frac{1}{s_{15}s_{34}}& & & &\\
&+\frac{1}{s_{34}s_{25}}& & & &\\
\\
-\frac{1}{s_{15}s_{34}}& -\frac{1}{s_{15}s_{34}}-\frac{1}{s_{34}s_{25}}&
\frac{1}{s_{13}s_{25}}+\frac{1}{s_{13}s_{24}} & \frac{1}{s_{15}s_{24}}+
\frac{1}{s_{15}s_{34}}&-\frac{1}{s_{15}s_{24}} &-\frac{1}{s_{13}s_{24}}
-\frac{1}{s_{15}s_{24}}\\
&&+\frac{1}{s_{15}s_{24}}+\frac{1}{s_{15}s_{34}}& & &\\
&&+\frac{1}{s_{34}s_{25}}\\
\\
-\frac{1}{s_{12}s_{34}}-\frac{1}{s_{15}s_{34}}&-\frac{1}{s_{15}s_{34}} &
\frac{1}{s_{15}s_{34}}+\frac{1}{s_{15}s_{24}} & \frac{1}{s_{12}s_{35}}+
\frac{1}{s_{12}s_{34}}&-\frac{1}{s_{15}s_{24}}-\frac{1}{s_{24}s_{35}} &
-\frac{1}{s_{15}s_{24}}\\
&&&+\frac{1}{s_{15}s_{34}}+\frac{1}{s_{15}s_{24}}&&\\
&&&+\frac{1}{s_{24}s_{35}}&&\\
\\
-\frac{1}{s_{15}s_{23}} & -\frac{1}{s_{14}s_{23}}-\frac{1}{s_{15}s_{23}}&
-\frac{1}{s_{15}s_{24}} &-\frac{1}{s_{15}s_{24}}-\frac{1}{s_{24}s_{35}} &
\frac{1}{s_{14}s_{35}}+\frac{1}{s_{14}s_{23}} &\frac{1}{s_{15}s_{23}}+\frac{1}{s_{15}s_{24}}\\
&&&&+\frac{1}{s_{15}s_{23}}+\frac{1}{s_{15}s_{24}}&\\
&&&&+\frac{1}{s_{24}s_{35}}&\\
\\
-\frac{1}{s_{23}s_{45}}-\frac{1}{s_{23}s_{15}}& -\frac{1}{s_{15}s_{23}}&
-\frac{1}{s_{13}s_{24}}-\frac{1}{s_{15}s_{24}}&-\frac{1}{s_{15}s_{24}} &
\frac{1}{s_{15}s_{23}}+\frac{1}{s_{15}s_{24}} &\frac{1}{s_{13}s_{45}}
+\frac{1}{s_{13}s_{24}}\\
&&&&&+\frac{1}{s_{15}s_{24}}+\frac{1}{s_{15}s_{23}}\\
&&&&&+\frac{1}{s_{23}s_{45}}
\end{pmatrix}.\nonumber\\&&\!\!\!\!\!\!\!\!\!\!\!
\eea
As in the four particle case,
we achieve color-kinematic symmetry by adding $\delta n$ to the Feynman rule determined numerators $n$ 
\bea
\bar n_i=n_i+\delta n_i \label{4.3},
\eea 
such that $\bar n$'s have the required symmetry.
 The net result is that we have the same set of equation 
as in (\ref{4.2}) with $n_i$'s replaced by $\delta n_i$ and with
each term with a $\Delta $ gaining a minus sign.  We now impose the requirement
that the Feynman numerator shifts by  $\delta n_i$ must leave the color ordered 
amplitudes untouched.  For example, we have
\bea
A(12345)&=&{n(12;3;45)\over s_{12}s_{45}}-{n(12;5;34)\over s_{12}s_{34}}
+{n(23;4;51)\over s_{23}s_{51}}\nonumber\\ &&-{n(23;1;45)\over s_{23}s_{45}}
-{n(34;2;51)\over s_{34}s_{51}}, \label{4.5}
\eea
which yields
\bea
&&{\delta n_1\over s_{12}s_{45}}+{\delta n_1-\delta n_{12}+\Delta (12|345)
\over s_{12}s_{34}}-{-\delta n_1+\delta n_{15}+\Delta (45|123)\over s_{23}s_{45}}
\nonumber\\ && +{\delta n_1-\delta n_{15}-\delta n_{14}+\delta n_6-\Delta (14|352)-\Delta (45|123)-\Delta (23|145)\over s_{23}s_{15}} \nonumber\\ &&
-{-\delta n_1+\delta n_{12}+\delta n_9-\delta n_6-\Delta (12|345)-\Delta (25|134)-\Delta (34|125)\over s_{34}s_{15}}=0,     \label{4.6} \eea
or, collecting all Jacobi-violating $\Delta$'s into a single quantity 
\bea
&& \delta n_1({1\over s_{12}s_{45}}+{1\over s_{12}s_{34}}
+{1\over s_{23}s_{15}}+{1\over s_{23}s_{45}}+{1\over s_{34}s_{15}})\nonumber\\& &
-\delta n_{12}({1\over s_{12}s_{34}}+{1\over s_{34}s_{15}})
+\delta n_{15}(-{1\over s_{23}s_{15}}-{1\over s_{23}s_{45}})\nonumber\\  &&
-\delta n_9 ({1\over s_{34}s_{15}})
+\delta n_{14}(-{1\over s_{23}s_{15}})
-\delta n_6 (-{1\over s_{23}s_{15}}-{1\over s_{34}s_{15}})
=D(12345), \label{4.7}
\eea
where 
\bea
D(12345)&\equiv& \Delta (12|345)(-{1\over s_{12}s_{34}}-{1\over s_{34}s_{15}})
+\Delta (45|123)({1\over s_{23}s_{15}}+{1\over s_{23}s_{45}})\nonumber\\ && 
+\Delta (14|352)({1\over s_{23}s_{15}})
+\Delta (23|145)({1\over s_{23}s_{15}})\nonumber \\ &&
+\Delta (25|134)(-{1\over s_{34}s_{15}})
+\Delta (34|125)(-{1\over s_{34}s_{15}}). \label{4.8}
\eea

In a similar fashion we obtain all the other $D$'s corresponding to the amplitudes in our chosen Kleiss-Kuijf basis, and we list them  in Appendix \ref{Appendix3}.

Succinctly, starting from the defining relation (\ref{4.9}),
\be
|A\rangle=M^{(5)}|\bar N\rangle,
\ee 
where $|A\rangle$ denotes the set of Kleiss-Kuijf amplitudes and $|\bar N\rangle$ the set of BCJ numerators, we replace $\bar n$'s by $n +\delta n$'s. 
On the other hand, the Feynman numerators 
$\langle N| =(n_1, n_6, n_9, n_{12}, n_{14}, n_{15}) $ satisfy
\be
|A\rangle-|D\rangle=M^{(5)}|N\rangle,
\ee
where we collected the Jacobi-violating terms into a six-component vector $|D\rangle$.
Then, the requirement for the shifts $\delta n$ is that they should satisfy  
\be
|D\rangle= M^{(5)} |\delta N\rangle,
\ee
or, more explicitly,
\be
\begin{pmatrix} D(12345)\\ D(14325) \\ D(13425) \\ 
D(12435) \\ D(14235) \\ D(13245) \end{pmatrix}=M^{(5)}
\begin{pmatrix} \delta n_1 \\ \delta n_6 \\ \delta n_9 \\  \delta n_{12} \\ \delta n_{14} \\ \delta  n_{15} \end{pmatrix}.
\label{4.10}
\ee
The solution for $\delta n_i$ is not unique, because 
$M^{(5)}$ has four eigenvectors with zero eigenvalue.  We gave a rather thorough discussion on this in \cite{VY3} with
regard to the origin of generalized gauge transformations.  The effects 
 are that we can determine only two linear combinations of 
$\delta n_i$, which are 
\bea
\delta n' &=&\delta n_1-\delta n_9{s_{12}s_{45}\over s_{13}s_{24}}
+\delta n_{12}{s_{45}(s_{12}+s_{24})\over s_{24}s_{35}}\nonumber\\ && 
-\delta n_{14}{s_{12}s_{45}\over s_{24}s_{35}}
+\delta n_{15}{s_{12}(s_{24}+s_{45})\over s_{13}s_{24}}\nonumber \\ &
=& s_{12}(s_{25}D(13425)-(s_{15}+s_{25})D(12435)), \label{4.11}
\eea
and
\bea
\delta n'' &=&\delta n_6+\delta n_9{s_{14}(s_{24}+s_{25})\over s_{13}s_{24}}
-\delta n_{12}{s_{14}s_{25}\over s_{24}s_{35}}\nonumber\\ && 
+\delta n_{14}{s_{25}(s_{14}+s_{24})\over s_{24}s_{35}}
-\delta n_{15}{s_{14}s_{25}\over s_{13}s_{24}}\nonumber\\ &
=&s_{25}(-(s_{12}+s_{15})D(13425)+s_{12}D(12435)). \label{4.12}
\eea
Another noteworthy remark is that they imply that there should be only 
two independent $D_i's$, which requires checking for consistency.

Using $D$'s and $\Delta$'s in Appendix \ref{Appendix3} and Appendix \ref{Appendix2}, respectively, we find that 
\be
\delta n'=s_{12}{s_{45}\over s_{24}}X, \ \ \ 
\delta n''=-s_{25}{s_{14}\over s_{24}}X,
\label{4.13}
\ee
where
\be
X={p_1^-\over p_1}(p_{52}-p_{54})
+{p_5^-\over p_5}(p_{12}-p_{14})-{p_3^-\over p_3}(p_{12}-p_{14}+
p_{52}-p_{54}),
\label{4.14}
\ee
and
\bea
&& p_{12}={p_1p_2\over p_1+p_4},\ \ p_{14}={p_1p_4\over p_1+p_2}, \nonumber\\ &&
p_{32}={p_3p_2\over p_3+p_4},\ \ p_{34}={p_3p_4\over p_3+p_2}, \nonumber\\  &&
p_{52}={p_5p_2\over p_5+p_4},\ \ p_{54}={p_5p_4\over p_5+p_2}. \label{4.15}
\eea
(We should attach a product of the five polarization tensors to 
$X$, which will be understood, because we are dealing with
on-shell amplitudes at this point.)

At this point it seems that the numerator shifts are bound to contain a large degree of ambiguity, since we are only placing a constraint on $\delta n'$ (\ref{4.11}) and on $\delta n''$ (\ref{4.12}). However, this is not the case if we impose the additional condition that the numerator shifts should not introduce spurious poles. For example, this would require that  
\bea
&&\delta n_1=s_{12}s_{45}a_1, \ \delta n_6=s_{14}s_{25}a_6, \ 
\delta n_{9}=s_{13}s_{25}a_{9}, \nonumber\\ && \delta n_{12}=s_{12}s_{35}a_{12}, \ 
\delta n_{14}=s_{14}s_{35}a_{14}, \ \delta n_{15}=s_{13}s_{45}a_{15}, \label{4.16}
\eea
where $a_1$ should have at most simple poles in $s_{12}$ and $s_{45}$, $a_6$ should have at most simple poles in $s_{14}$ or $s_{25}$ etc.

We take note that $X$ is symmetric under $1\leftrightarrow 5$, 
but antisymmetric under $2\leftrightarrow 4$. Then (\ref{4.13}) can be written as
\be
s_{24}a_1-s_{25}a_{9}+(s_{12}+s_{24})a_{12}
-s_{14}a_{14}+(s_{24}+s_{45})a_{15}=X, \label{4.17}
\ee
and
\be
s_{24}a_6+(s_{24}+s_{25})a_9-s_{12}a_{12}
+(s_{14}+s_{24})a_{14}-s_{45}a_{15}=-X. \label{4.18}
\ee
Actually (\ref{4.17}) and (\ref{4.18}) follow from each other, because
under
\be
1\leftrightarrow 5: \ \  a_1 \leftrightarrow -a_6, \
a_{12} \leftrightarrow -a_9, \ a_{15} \leftrightarrow -a_{14},  \label{4.19}
\ee
and under
\be
2\leftrightarrow 4: \ \  a_1\leftrightarrow a_6, \
a_{15} \leftrightarrow a_9, \  a_{12} \leftrightarrow a_{14}.  \label{4.20}
\ee
When we add $(\ref{4.17}-\ref{4.18})$, we further obtain
\be
a_1+a_6+a_9+a_{12}+a_{14}+a_{15}=0.  \label{4.21}
\ee
To solve for $\delta n_i$, or equivalently for the $a_i$ separately, instead of just the combinations 
$\delta n'$ and $\delta n''$, we are guided by symmetry and by the requirement that $a$'s must have at most simple poles in the allowed channels: e.g. $a_1$ can have at most simple poles in $s_{12}$ and in $s_{45}$ etc. 
\bea
a_1&=&\ \ \ {p_1^-\over p_1}\big(x_1^{32}p_{32}+x_1^{34}p_{34}
+x_1^{52}p_{52}+x_1^{54}p_{54}\big)\nonumber\\ &&
+{p_3^-\over p_3}\big(y_1^{12}p_{12}+y_1^{14}p_{14}
+y_1^{52}p_{52}+y_1^{54}p_{54}\big)\nonumber\\ &&
+{p_5^-\over p_5}\big(z_1^{12}p_{12}+z_1^{14}p_{14}
+z_1^{32}p_{32}+z_1^{34}p_{34}\big),  \label{4.22}
\eea
\bea
a_{15}&=& {p_1^-\over p_1}\big(x_{15}^{32}p_{32}+x_{15}^{34}p_{34}
+x_{15}^{52}p_{52}+x_{15}^{54}p_{54}\big)\nonumber\\ &&
+{p_3^-\over p_3}\big(y_{15}^{12}p_{12}+y_{15}^{14}p_{14}
+y_{15}^{52}p_{52}+y_{15}^{54}p_{54}\big)\nonumber\\ &&
+{p_5^-\over p_5}\big(z_{15}^{12}p_{12}+z_{15}^{14}p_{14}
+z_{15}^{32}p_{32}+z_{15}^{34}p_{34}\big),  \label{4.23}
\eea
where the $x$'s, $y$'s, and $z$'s are functions of $s_{ij}$. By inspection, from (\ref{4.13}) and (\ref{4.14}) we infer that they are of the order $1/s$.  We 
obtain the other $a$'s through (\ref{4.19}-\ref{4.20}). 
After some straightforward but tedious algebra, recorded in Appendix \ref{Appendix4}, we obtain
\bea
&& x_1^{32}={1\over s_{12}}+\frac{\alpha}{s_{45}}, \ \ x_1^{34}=\frac{\beta}{s_{45}}, \ \ x_1^{52}=-{1\over s_{12}}+\frac{\beta}{s_{45}}, \ \ x_1^{54}=\frac{\alpha}{s_{45}}, \nonumber\\ &&
y_1^{12}=\alpha\bigg(\frac{1}{s_{12}}-\frac{1}{s_{45}}\bigg), \ \ y_1^{14}=\beta\bigg(\frac{1}{s_{12}}-\frac{1}{s_{45}}\bigg), \ \ y_1^{52}=\beta\bigg(\frac{1}{s_{12}}-\frac{1}{s_{45}}\bigg), \ \ y_1^{54}=\alpha\bigg(\frac{1}{s_{12}}-\frac{1}{s_{45}}\bigg), \nonumber\\&&
z_1^{12}=-\frac{\alpha}{s_{12}}, \ \ z_1^{14}={1\over s_{45}}-\frac{\beta}{s_{45}}, \ \ z_1^{32}=-\frac{\beta}{s_{12}}, 
\ \ z_1^{34}=-{1\over s_{45}}-\frac{\alpha}{s_{12}}, \label{4.50}\\
&& x_{15}^{32}={1\over s_{13}}-\frac{\alpha}{s_{45}}, \ \ x_{15}^{34}=-{1\over s_{13}}-\frac{\beta}{s_{45}}, \ \ x_{15}^{52}=-\frac{\beta}{s_{45}}, \ \ x_{15}^{54}=-\frac{\alpha}{s_{45}}, \nonumber\\ &&
y_{15}^{12}=-{1\over s_{13}}+\frac{\alpha}{s_{45}}, \ \ y_{15}^{14}={1\over s_{13}}+\frac{\beta}{s_{45}}, \ \ y_{15}^{52}=\frac{\beta}{s_{45}}, \ \ y_{15}^{54}=\frac{\alpha}{s_{45}}, \nonumber\\ &&
z_{15}^{12}=\frac{\beta-\alpha}{s_{13}}, \ \ z_{15}^{14}=-{1\over s_{45}}+\frac{\alpha-\beta}{s_{13}}, \ \ z_{15}^{32}=\frac{\alpha-\beta}{s_{13}}, 
\ \ z_{15}^{34}=\frac{\beta-\alpha}{s_{13}}+{1\over s_{45}}. \label{4.51}
\eea

Please be reminded that $\delta n_1=s_{12}s_{45}a_1$ and 
$\delta n_{15}=s_{13}s_{45}a_{15}$.  Thus, there is no spurious 
singularity in the forms of ${1\over s_{12}}, \ {1\over s_{13}}$
or ${1\over s_{45}}$ in $\delta n_1$ or $\delta n_{15}$, nor is there
any in other $\delta n$'s.  
The numerator shifts are not uniquely determined, there is still some arbitrariness as parametrized by the constants $\alpha$ and $\beta$. This is due to the fact that we have the freedom of shifting the numerators using the zero-modes of the propagator matrix. This freedom was further restricted here by requiring that the shifts preserve the original pole structure of the Feynman-rules amplitude decomposition (\ref{1.1}), leaving only the undetermined $\alpha$ and $\beta$.

Also, we would like to point out that
if we choose $|+\rangle[+|\propto P_2$ or $|+\rangle[+|\propto P_4$, which makes $\epsilon ^- _2=0$ 
or $\epsilon ^-_4 =0$, respectively, then the shifts $\delta n_i =0$,
or the numerators are already BCJ symmetric to begin with\footnote{
For MHV amplitudes,  space-cone gauge with $|+\rangle[+|\propto P_i$, where $i$ denotes an on-shell negative helicity gluon, yields BCJ numerators. The other external legs can be kept off-shell.
The choice made such that one of negative helicity gluons is reference (and the space-cone gauge is defined relative to it) means that the vertices used to generate the MHV diagrams will be of type $(++-)$ and only one $(+-\;-)$, where one of the negative helicity gluons participating in the only vertex $(+-\;-)$ is our reference gluon. The quartic vertex $(++-\;-)$ is zero provided that we make this choice. With this structure one can easily check that the numerators generated by Feynman rules are color-kinematic symmetric. See also \cite{Monteiro:2011pc}.
}.

\section { Effective Lagrangian:}\label{section5}

Using the results of (\ref{4.50}-\ref{4.51}), we can derive all the 
\be
a(ij;k;lm)\equiv {\delta n(ij;k;lm)\over s_{ij}s_{lm}}, \label{5.1}
\ee
which appear naturally in the scattering amplitudes, with an effective Lagrangian
\bea
{\cal L}_5&=&\bigg[
-\frac{f_{c_1 c_2 d}}{s_{c_1 c_2}}
 (f_{d c_3 e} f_{e c_4 c_5} +  f_{d c_5 e} f_{e c_3 c_4}+ f_{d c_4 e} f_{e c_5 c_3} ) \nonumber\\
&&\!\!\!\!\!\!\!\!\!\! -\frac{f_{c_1 c_5 d}}{s_{c_1c_5}}
(f_{d c_2 e} f_{c_3 c_4 e} + f_{d c_3 e} f_{c_4 c_2 e} + f_{d c_4 e} f_{c_2 c_3 e})
\nonumber\\
&&\!\!\!\!\!\!\!\!\!\! - (1-\alpha+\beta)\frac{f_{c_2 c_4 d}}{s_{c_2 c_4}}
(f_{d c_1 e} f_{c_3 c_5 e} + f_{d c_3 e} f_{c_5 c_1 e}+f_{d c_5 e} f_{c_1 c_3 e})
\nonumber\\
&&\!\!\!\!\!\!\!\!\!\!+ \alpha\bigg(
 \frac{f_{c_5 c_2 d}}{s_{c_2 c_5}}
 ( f_{d c_1 e} f_{e c_4 c_3} + f_{d c_4 e} f_{e c_3 c_1}+ f_{d c_3 e} f_{e c_1 c_4})
+\frac{f_{c_3 c_4 d}}{s_{c_3 c_4}}
(f_{d c_1 e} f_{e c_2 c_5}+f_{d c_2 e} f_{e c_5 c_1}+f_{d c_5 e} f_{e c_1 c_2})\bigg)\nonumber\\
&&\!\!\!\!\!\!\!\!\!\!+\beta \bigg(\frac{f_{c_5 c_4 d}}{s_{c_4 c_5}}
(f_{d c_1 e} f_{e c_2 c_3} + f_{d c_2 e} f_{e c_3 c_1} + f_{d c_3 e} f_{e c_1 c_2})
+\frac{f_{c_3  c_2 d}}{s_{c_3 c_2}} 
(f_{d c_1 e}f_{e c_4 c_5}+ f_{d c_4 e} f_{e c_5 c_1}+f_{d c_5 e}f_{e c_1 c_4})\bigg)
\nonumber\\
&&\!\!\!\!\!\!\!\!\!\!+(\beta-\alpha)\frac{f_{c_5 c_3 d}}{s_{c_3 c_5}}(f_{d c_2 e} f_{e c_4 c_1}+f_{d c_4 e}f_{e c_1 c_2}
+ f_{d c_1 e}f_{e c_2 c_4})\bigg]\frac{\partial^-}{\partial} a_{c_1}^+ \partial a_{c_2}^- \frac{1}{\partial}(\partial a_{c_5}^+ a_{c_4}^-) a_{c_3}^+ + h.c.
\eea
where $\alpha$ and $\beta$ are arbitrary constants. 
We should note that in view of the Jacobi identity obeyed by the structure constants $f_{abc}$ this effective Lagranigan is null,
which is of course a succinct statement that the shifts we performed in the numerators  have no effects on the physical amplitudes.

It is easy to check that this effective Lagrangian implements the desired shifts:
\bea
a_1&=& a(1^+2^-;3^+;4^-5^+)= \frac{p_1^-}{p_1}\bigg[\frac{1}{s_{12}}(p_{32}-p_{52})+
\frac{1}{s_{45}}\bigg( \alpha ( p_{32}+p_{54}) + \beta(p_{34}+p_{52})\bigg)\bigg]\nonumber\\
&&\!\!\!\!\!\!\!\!\!\!\!
+\frac{p_3^-}{p_3}\bigg[\frac{1}{s_{12}}\bigg(\alpha(p_{12}+p_{54})+\beta(p_{14}+p_{52})\bigg)
-\frac{1}{s_{45}}\bigg(\alpha(p_{12}+p_{54})+\beta(p_{14}+p_{52})\bigg)\bigg]\nonumber\\
&&\!\!\!\!\!\!\!\!\!\!\!
+\frac{p_5^-}{p_5}\bigg[-\frac{1}{s_{12}}\bigg(\alpha(p_{12}+p_{34})+\beta(p_{14}+p_{32})\bigg)
+\frac{1}{s_{45}}(p_{14}-p_{34})\bigg]
\eea
and 
\bea
a_{15}&=&a(1^+3^+;2^-;4^-5^+)=\frac{p_1^-}{p_1}\bigg[\frac{1}{s_{13}}(p_{32}-p_{34})-
\frac{1}{s_{45}}\bigg(\alpha(p_{32}+p_{54})+\beta(p_{34}+p_{52})\bigg)\bigg]
\nonumber\\
&&\!\!\!\!\!\!\!\!\!\!\!
+\frac{p_3^-}{p_3}\bigg[\frac{1}{s_{13}}(p_{14}-p_{12})+
\frac{1}{s_{45}}\bigg(\alpha(p_{12}+p_{54})+\beta(p_{14}+p_{52})\bigg)\bigg]\nonumber\\
&&\!\!\!\!\!\!\!\!\!\!\!
+\frac{p_5^-}{p_5}\bigg[\frac{\beta-\alpha}{s_{13}}(p_{12}-p_{14}-p_{32}+p_{34}) + \frac{1}{s_{45}}(p_{34}-p_{14})\bigg]
\eea
We will not write  out all the $a(ij;k;lm)$ explicitly, as their particular form is not especially illuminating.  Suffices it to say that they fall into three groups, according to their helicity arrangements:
\be
(A) \qquad  \ \ \ (\pm \mp;+; \pm \mp),
\ee
which consists of
\be
a(12;5;34), \ a(12;3;45), \ a(14;5;23), \ a(14;3;52), \ a(23;1;45), 
\ a(25;1;34). \label{5.3}
\ee
The $\alpha=\beta=0$ contribution to the set of numerators (A) comes from the piece of 
${\cal L}_{5}$ which is proportional to 
\be
(f_{dec_3}f_{c_4c_5e}
+f_{dec_5}f_{c_3c_4e})f_{c_2c_1d}{1\over s_{c_1 c_2}}\label{5.4}
\ee
\be (B)\qquad  \ \ \ (\pm \mp;-; ++),\ \ (++;-;\pm \mp),
\ee
which consists of
\be
a(12;4;53),\ a(13;2;45),\ a(13;4;52),\ a(14;2;35),\ a(15;2;34),
\ a(15;4;23). \label{5.5}
\ee
The $\alpha=\beta=0$ contribution to this set of numerators comes from terms proportional to 
\be
f_{dec_4}f_{c_5c_3e}
f_{c_2c_1d}{1\over s_{c_1 c_2}}
+(f_{dec_2}f_{c_3c_4e}
+f_{dec_4}f_{c_2c_3e})f_{c_5c_1d}{1\over s_{c_1 c_5}}\label{5.6}
\ee
\be 
(C) \qquad  \ \ \ (++;+; --), \ \ (--;+;++),
\ee
which consists of
\be
 a(13;5;24),\  a(15;3;42), \ a(24;1;35). \label{5.7}
\ee
Lastly, the $\alpha=\beta=0$ contribution to the numerators of type (C) comes from terms proportional to 
\be
f_{dec_3}f_{c_4c_2e}f_{c_5c_1d}{1\over s_{c_1 c_5}}
+ (f_{dec_1}f_{c_3c_5e}+f_{dec_3}f_{c_5c_1e}+f_{dec_5}f_{c_1c_3e})
f_{c_4c_2d}{1\over s_{c_2c_4}}. \label{5.8}
\ee

\section{Concluding Remarks} \label{section6}

We would like to digress at this point and to explain how
BCFW on-shell recursion \cite{BCFW} is performed in the space-cone gauge.  
We note
that the Lagrangian in (\ref{2.6}) has no $\bar \partial $
dependence in its interaction terms.  Thus, analytical 
continuation is done by making shifts in some $\bar p$
direction with a complex number z (and if necessary by
also choosing some appropriate reference vector $\pm $
so that $A(z)\to 0$ as $z \to \infty$.)  Because the 
numerators have no $\bar p$ dependence, the continuation
does not affect them and the poles of the amplitude are
due to the vanishing of some inverse propagators.  This
polology makes it very transparent the meaning of cuts of the
amplitudes in evaluating the integral $\int {dz\over z} A(z)$.
In other words, the cutting of the amplitude into two halves
gives an easy organization to yield BCFW recurrence \cite{VY1}.

A question which can be asked is whether one can circumvent the Lagrangian approach and write down
BCJ numerators from amplitudes.
In particular, as noticed in \cite{VY2}, there is a set of BCJ numerators which can be obtained from knowledge of the amplitudes provided that we use the zero modes of the propagator matrix $M$ fully to set to zero $(n-3)(n-3)!$ components of the BCJ numerators $\bar N$. Then the relation $|A\rangle = M|\bar N\rangle$  can be inverted. However, the numerators obtained in such fashion will generally contain spurious poles. In \cite{Naculich:2014rta}, with the same starting point, it was noticed that one can obtain `virtuous'' numerators by applying a certain symmetrization procedure.  While these expression  carry
certain 'virtues' \cite{Broedel:2011pd}, there are also issues which demand
attention.  Consider the virtuous four-point numerator given by \cite{Broedel:2011pd,Naculich:2014rta}
\be
\hat n (1^-2^+;3^+4^-)={1\over 3}
(s_{12}A(1^-2^+3^+4^-)-s_{14}A(1^-4^-2^+3^+)).
\ee
When we use 
\be
A(1^-4^-2^+3^+)=A(4^-2^+3^+1^-)
=A(1^-2^+3^+4^-)|_{1 \leftrightarrow 4}
\ee
and (\ref{3.26}), we express it as
\bea
 \hat n (1^-2^+;3^+4^-)
&=& {1\over 3}p_1p_2p_3p_4\bigg({p_2^-\over p_2}-{p_3^-\over p_3}\bigg)^2
\bigg({p_4^+\over p_4}-{p_1^+\over p_1}\bigg)^2
(-{1\over s_{14}}+{1\over s_{24}}) 
\epsilon(1)^-\epsilon(2)^+\epsilon(3)^+\epsilon(4)^-.\nonumber\\
\eea
We see that the numerators obtained in this way contain spurious poles at $s_{14} $ and $s_{24}$. This defeats to some extent the purpose of decomposing the amplitude in the form (\ref{1.1}), with the propagator poles manifestly written. 

Color-kinematics duality allows for a particular version of the KLT relations, expressing the gravity amplitudes in terms of gauge theory amplitudes, with the key ingredient being the color-kinematic symmetric (BCJ) gauge theory numerators.  However, the BCJ symmetry
is not automatic, if one is to compute the numerators from a
gauge Lagrangian.  To summarize our results, what we have shown is that the violation of
this symmetry can be systematically computed
 and absorbed
into shifts of the Feynman numerators. These shifts do not change the color-ordered
amplitudes.  We specifically work in a light-like
gauge, because it is physical and therefore makes the on-shell limit
transparent.  We have set up a set of equations for four and five
particle cases, which  are used to solve for the shifts in terms of
the violations.  In the four particle case, there is no need to
make any shift on-shell.  For the five particle case, we have
obtained the general solution for shifts which are consistent
with the acceptable pole structure.  We have also constructed the null five-point Lagrangian which augments the light-like gauge fixed Lagrangian and which yields color-kinematic symmetric numerators. It is clear
that this program should work for any number of particles and
with an arbitrary choice of the light-like gauge fixing vector. 


\section*{Acknowledgments}

This work was supported, in part, by the U.S. Department
of Energy under Grant No.~DE-SC0007984.

\appendix
\section{ Four-point off-shell numerators}\label{Appendix1}

In this appendix, we calculate the Jacobi-permutation of
three indices of four particle numerators.

 It is useful to notice that for a tree-level amplitude, the net factor of 2 from the propagators $i/(2 s_{ij})$ and from the vertices given in Section \ref{section2} will cancel against the normalization factor of the group generators. [See footnote 6.] 
In what follows we decompose the amplitude as in (\ref{1.1}) and (\ref{tree color}).  The factors of $g^{n-2}$  and $(-i)$ which accompany a tree-level $n$-point amplitude are implicit. We will omit them in writing out the numerators of a color-ordered amplitude.
With this observation, we have the following ingredients for the color-ordered amplitudes: the propagator is $1/P^2$, the color-ordered three-point vertex is
\bea
{\rm{ 3\,pt \,vertex}}(1^-2^+k^-)&\equiv&(1^-2^+k^-)=
  p_2(\frac{k^+}{k}-\frac{p_1^+}{p_1}),\nn\\
{\rm{3\,pt\,vertex}}(1^-2^+k^+)&\equiv&(1^-2^+k^+)=p_1 (\frac{p_2^-}{p_2}-
\frac{k^-}{k}),
\eea
 and  color-ordered four-point vertex is 
\bea
{\rm{ 4\,pt\,vertex}}(1^-2^+;3^+4^-)&=&-\frac{p_2p_4+p_1p_3}{(p_1+p_2)^2}\nn\\
{\rm{ 4\,pt\,vertex}}(4^-1^-;2^+3^+)&=&0\nn\\
{\rm{ 4\,pt\,vertex}}(2^+1^-;3^+4^-)&=&{\rm{ 4\,pt\,vertex}}(1^-2^+;4^-3^+)\nn\\
&=&
-{\rm{ 4\,pt\,vertex}}(1^-2^+;3^+4^-)=-{\rm{ 4\,pt\,vertex}}(2^+1^-;4^-3^+).\nn\\
\eea
 Of course, the color-ordered ${\rm{ 4\,pt\,vertex}}(1234)$ is the sum of ${\rm{ 4\,pt\,vertex}}(12;34)$ and ${\rm{ 4\,pt\,vertex}}(41;23)$. The split we make is relevant only in assigning each contribution to a certain numerator: ${\rm{ 4\,pt\,vertex}}(12;34)$ times the inverse propagator $-s_{12}$ contributes to $n(12;34)$ and ${\rm{ 4\,pt\,vertex}}(41;23)$ times the inverse propagator $-s_{14}$ contributes to $n(41;23)$.

Schematically, we write
\be
n(12;34)=(12k)(k34)+(12;34),
\ee
 where
\be
(12;34)\equiv{\rm{ 4\,pt\,vertex}}(12;34) \times (-s_{12})
\ee now includes the inverse propagator.\footnote{To avoid cluttering the notation further we write $(12k)(k34)$ for the cubic vertex contribution even though
we mean that in each vertex all momenta are incoming, and so this should really be written as $(12k)(-k34)$, with $P_k=-P_1-P_2=P_3+P_4$. We hope that this is an obvious omission and will refrain from writing the sign of the momentum in the other cubic vertex.}
Thus, with the understanding that a
product of the four polarization tensors is omitted and that the 
particles can then be off-shell, we have
\bea
n(1^-2^+;3^+4^-)&=&
({p_2^-\over p_2}-{p_1^-+p_2^-\over p_1+p_2})p_1
({p_4^+\over p_4}-{p_1^++p_2^+\over p_1+p_2})p_3\nonumber\\ &&
+({p_1^++p_2^+\over p_1+p_2}-{p_1^+\over p_1})p_2
({p_1^-+p_2^-\over p_1+p_2}-{p_3^-\over p_3})p_4\nonumber\\ &&
+{p_2p_4+p_1p_3\over(p_1+p_2)^2}  \tfrac 12 (P_1+P_2)^2,\label{A1-1}
\eea
where the last term is $(12;34)$, and as explained before it includes the  $\tfrac 12(P_1+P_2)^2$ factor.    We pick out one term each from the three lines above to form
\be
{p_2p_4+p_1p_3\over (p_1+p_2)^2}[(p_1^++p_2^+)(p_1^-+p_2^-)
+ \tfrac 12 (P_1+P_2)^2]={p_2p_4+p_1p_3\over (p_1+p_2)^2}(p_1+p_2)
(\bar p_1+\bar p_2). \label{A1-2}
\ee
Now we write
\be
\bar p_1+\bar p_2=-\bar p_3-\bar p_4
=-({p_3^+p_3^-\over p_3}+{ \tfrac 12 P_3^2 \over p_3})
-({p_4^+p_4^-\over p_4}+{ \tfrac 12 P_4^2 \over p_4}).\label{A1-3}
\ee
Putting (\ref{A1-1},\ref{A1-2}) into (\ref{A1-3}), we have
\bea
n(1^-2^+;3^+4^-)&=&
p_1p_3\bigg({p_3^-\over p_1+p_2}{p_4^+\over p_4}
+{p_2^-\over p_2}{p_4^+\over p_4} 
+{p_2^-\over p_2}{p_3^++p_4^+\over p_1+p_2}\nonumber\\ &&
\ \ \ -{1\over p_1+p_2}{p_3^+p_3^-\over p_3}
-{1\over p_1+p_2}{ \tfrac 12 P_3^2\over p_3}
-{1\over p_1+p_2}{ \tfrac 12 P_4^2\over p_4}\bigg)\nonumber\\ &&
+p_2p_4\bigg({p_4^+\over p_1+p_2}{p_3^-\over p_3}
+{p_1^+\over p_1}{p_3^-\over p_3} 
+{p_1^+\over p_1}{p_3^-+p_4^-\over p_1+p_2}\nonumber\\ &&
\ \ \ -{1\over p_1+p_2}{p_4^+p_4^-\over p_4}
-{1\over p_1+p_2}{\tfrac 12 P _3^2\over p_3}
-{1\over p_1+p_2}{ \tfrac 12 P_4^2\over p_4}\bigg).\label{A1-4}
\eea

It is useful to add and subtract $-(p_1p_3+p_2p_4)
({p_2^-\over p_2}{p_4^+\over p_4}-{p_2^-\over p_2}{p_1^+\over p_1}
-{p_3^-\over p_3}{p_4^+\over p_4}+{p_3^-\over p_3}{p_1^+\over p_1})$
to the expression above.  Then we use 
\be
p_1p_3({p_3^-\over p_1+p_2}{p_4^+\over p_4}+
{p_3^-\over p_3}{p_4^+\over p_4})
+p_2p_4({p_4^+\over p_1+p_2}{p_3^-\over p_3}+
{p_3^-\over p_3}{p_4^+\over p_4})=-p_3^-p_4^+,\label{A1-5}
\ee
and 
\bea
&&p_1p_3({p_2^-\over p_2}{p_3^++p_4^+\over p_1+p_2}
 +{p_2^-\over p_2}{p_1^+\over p_1})
+p_2p_4({p_1^-\over p_1}{p_3^-+p_4^-\over p_1+p_2}
+{p_2^-\over p_2}{p_1^+\over p_1})\nonumber\\ &
=&-{p_1p_3\over p_1+p_2}{p_2^+p_2^-\over p_2}
-{p_2p_4\over p_1+p_2}{p_3^+p_3^-\over p_3}-p_1^+p_2^-,
\label{A1-6}
\eea 
to obtain
\bea
n(1^-2^+;3^+4^-)  & =& - {1\over p_1+p_2} 
\bigg((p_1p_3+p_2p_4)({ P_3^2\over 2 p_3}+{ P_4^2\over2 p_4})\nonumber\\ &&
+p_1p_3({p_2^+p_2^-\over p_2}+{p_3^+p_3^-\over p_3})
+p_2p_4({p_1^+p_1^-\over p_1}+{p_4^+p_4^-\over p_4})\bigg)\nonumber\\ &&
+(p_1p_3+p_2p_4)
({p_2^-\over p_2}{p_4^+\over p_4}-{p_2^-\over p_2}{p_1^+\over p_1}
-{p_3^-\over p_3}{p_4^+\over p_4}-{p_3^-\over p_3}{p_1^+\over p_1})
\nonumber\\ &&
=s_{23}+({p_2^-\over p_2}-{p_3^-\over p_3})
({p_4^+\over p_4}-{p_1^+\over p_1})(p_1 p_3+p_2 p_4) \nonumber\\ &&
+{p_1p_3\over2( p_1+p_2)}(-{P_4^2\over p_4}+{P_2^2\over p_2})
+{p_2p_4\over2( p_1+p_2)}(-{P_3^2\over p_3}+{P_1^2\over p_1}).
\label{A1-7}
\eea

In a similar way, we obtain $n(3^+1^-;2^+4^-)$ and 
$n(2^+3^+;1^-4^-)$ given in (\ref{3.20}-\ref{3.21}).

\section{Five-point $\Delta$'s}\label{Appendix2}

In this appendix we calculate the $\Delta $'s for $1^+2^-3^+4^-5^+$.  We 
put all the external particles on-shell.  To shorten the expression of various terms we continue to omit
the common factor of the product of the polarizations (i.e. the external line factors).  Let us take one specific case and the others will be treated similarly.  For 
$\Delta (1^+2^-|3^+4^-5^+)$,
there are two sets of contributions.  The first set is due to a four-vertex
multiplied by a three vertex for each graph.  The second set is 
due to a three vertex $(1^+, 2^-, -(1+2)^+)$ multiplied by the
off-shell $\Delta ( (1+2)^-|3^+4^-5^+).$  
For the first set, we have
\bea
\Delta (1^+2^-|3^+4^-5^+)_1&=& (1^+2^-;3^+ (4+5)^-)(-(4+5)^+4^-5^+)\nonumber\\ &&
+(1^+2^-;5^+ (3+4)^-)(-(3+4)^+3^+4^-)\nonumber\\&&
+(1^+2^-;4^+(5+3)^+)(-(5+3)^-5^+3^+) \nonumber\\&&
={s_{12}\over (p_1+p_2)^2}\big[
(p_3p_2+p_1(p_4+p_5))
({p_5^-\over p_5}-{p_4^-+p_5^-\over p_4+p_5})p_4\nonumber\\&&
+(p_5p_2+p_1(p_3+p_4))
({p_3^-+p_4^-\over p_3+p_4}-{p_3^-\over p_3})p_4 \nonumber\\ &&
+(p_2(p_3+p_5))+p_1p_4)({p_5^-\over p_5}-{p_3^-\over p_3})(p_3+p_5)
\big],
\label{A2-1}
\eea
which after some algebra is simplified to
\bea
\Delta (1^+2^-|3^+4^-5^+)_1&=&
{s_{12}\over (p_1+p_2)}
\big[-{p_5^-\over p_5}(p_2p_3+p_1p_4)
+{p_3^-\over p_3}(p_1p_4+p_2p_5)\nonumber\\ &&
+{p_3^-+p_4^-\over p_3+p_4}p_2p_3
-{p_4^-+p_5^-\over p_4+p_5}p_2p_5\big].\label{A2-2}
\eea
The contribution of the other set is
\bea
\Delta (1^+2^-|3^+4^-5^+)_2&=&
(1^+, 2^-, -(1+2)^+)\Delta ( (1+2)^-/3^+4^-5^+)\nonumber\\ &
=&s_{12}({p_1^-+p_2^-\over p_1+p_2}-{p_1^-\over p_1})p_2
p_4({1\over p_4+p_5}-{1\over p_3+p_4}). \label{A2-3}
\eea
The sum of these two contributions gives
\bea
\Delta (1^+2^-|3^+4^-5^+)&=&s_{12}\big[                             
{p_1^-\over p_1}p_2p_4(-{1\over p_4+p_5}+{1\over p_3+p_4})\nonumber\\ &&
+{p_3^-\over p_3}({p_1p_4\over p_1+p_2}-{p_2p_5\over p_4+p_5})\nonumber\\ &&
+{p_5^-\over p_5}({p_2p_3\over p_3+p_4}-{p_1p_4\over p_1+p_2})
\big].\label{A2-4}
\eea
Please note that 
\be
\Delta (ij|klm)=-\Delta (klm|ij)=-\Delta (ji|klm)=-\Delta (ij|lkm)  =-\Delta (mlk|ji),
 \label{A2-5}
\ee
and therefore, we have 
\bea
\Delta (1^+2^-3^+|4^-5^+)&=&-\Delta (5^+4^-|3^+2^-1^+)
=-\Delta (1^+2^-|3^+4^-5^+)|_{1\leftrightarrow 5, 2\leftrightarrow 4}
\nonumber\\&=&-s_{45}\big[ 
{p_1^-\over p_1}({p_3p_4\over p_2+p_3}-{p_2p_5\over p_4+p_5})\nonumber\\ &&
 +{p_3^-\over p_3}({p_2p_5\over p_4+p_5}-{p_1p_4\over p_1+p_2})
\nonumber\\ &&
+{p_5^-\over p_5}p_2p_4(-{1\over p_1+p_2}+{1\over p_2+p_3})
\big].\label{A2-6}
\eea
In a similar fashion, we obtain
\bea
\Delta (1^+4^-5^+|2^-3^+)&=&\Delta (1^+2^-3^+|4^-5^+)|_
{2\leftrightarrow 4, 1\leftrightarrow 5}
\nonumber\\ &=&-s_{23}\big[ 
{p_1^-\over p_1}({p_2p_5\over p_4+p_5}-{p_3p_4\over p_2+p_3})\nonumber\\ &&
+{p_5^-\over p_5}({p_3p_4\over p_2+p_3}-{p_1p_2\over p_1+p_4})
\nonumber\\ &&
+{p_3^-\over p_3}p_2p_4(-{1\over p_1+p_4}+{1\over p_4+p_5})
\big],
\label{A2-7}
\eea
\bea
\Delta (1^+2^-5^+|3^+4^-)&=&-\Delta (1^+2^-3^+|4^-5^+)|_
{3\leftrightarrow 5}
\nonumber\\ &=&s_{34}\big[ 
{p_1^-\over p_1}({p_4p_5\over p_2+p_5}-{p_2p_3\over p_3+p_4})\nonumber\\ &&
+{p_5^-\over p_5}({p_2p_3\over p_3+p_4}-{p_1p_4\over p_1+p_2})
\nonumber\\ &&
+{p_3^-\over p_3}p_2p_4(-{1\over p_1+p_2}+{1\over p_2+p_5})
\big],\label{A2-8}
\eea
\bea
\Delta (3^+4^-5^+|1^+2^-)&=&\Delta (1^+2^-5^+|3^+4^-)|_
{2\leftrightarrow 4, 1\leftrightarrow 3}
\nonumber\\ &=&s_{12}\big[ 
{p_3^-\over p_3}({p_2p_5\over p_4+p_5}-{p_1p_4\over p_1+p_2})\nonumber\\ &&
+{p_5^-\over p_5}({p_1p_4\over p_1+p_2}-{p_2p_3\over p_3+p_4})\nonumber\\ &&
+{p_1^-\over p_1}p_2p_4(-{1\over p_3+p_4}+{1\over p_4+p_5})
\big],\label{A2-9}\eea
\bea
\Delta (1^+3^+4^-|2^-5^+)&=&-\Delta (3^+4^-5^+|1^+2^-)|_
{ 1\leftrightarrow 5}
\nonumber\\ &=&-s_{25}\big[ 
{p_3^-\over p_3}({p_1p_2\over p_1+p_4}-{p_4p_5\over p_2+p_5})\nonumber\\ &&
+{p_1^-\over p_1}({p_4p_5\over p_2+p_5}-{p_2p_3\over p_3+p_4})\nonumber\\ &&
+{p_5^-\over p_5}p_2p_4(-{1\over p_3+p_4}+{1\over p_1+p_4})
\big],\label{A2-10}\eea
\bea
\Delta (2^-3^+5^+|1^+4^-)&=&-\Delta (1^+2^-3^+|4^-5^+)|_
{ 1\leftrightarrow 5}
\nonumber\\ &=&-s_{14}\big[ 
{p_3^-\over p_3}({p_4p_5\over p_2+p_5}-{p_1p_2\over p_1+p_4})\nonumber\\ &&
+{p_5^-\over p_5}({p_1p_2\over p_1+p_4}-{p_3p_4\over p_2+p_3})\nonumber\\ &&
+{p_1^-\over p_1}p_2p_4(-{1\over p_2+p_3}+{1\over p_2+p_5})
\big],\label{A2-11}\eea
\bea
\Delta (1^+2^-4^-|3^+5^+)&=&-s_{35}\big[ p_1(p_3+p_5)({p_3^-\over p_3}
-{p_5^-\over p_5})({1\over p_1+p_4}-{1\over p_1+p_2})\big], 
\label{A2-12}\eea
\bea
\Delta (2^-3^+4^-|1^+5^+)&=&
-\Delta (1^+2^-4^-|3^+5^+)|_{1\leftrightarrow 3}\nonumber\\ &
=&s_{15}\big[ p_3(p_1+p_5)({p_5^-\over p_5}
-{p_1^-\over p_1})({1\over p_2+p_3}-{1\over p_3+p_4})\big],\label{A2-13}\eea
\bea
\Delta (2^-4^-5^+|1^+3^+)&=&
-\Delta (1^+2^-4^-/3^+5^+)|_{1\leftrightarrow 5}\nonumber\\ &
=&s_{13}\big[ p_5(p_1+p_3)({p_3^-\over p_3}
-{p_1^-\over p_1})({1\over p_4+p_5}-{1\over p_2+p_5})\big], 
\label{A2-14}
\eea
and 
\be
\Delta (1^+3^+5^+|2^-4^-)=0.  \label{A2-15}
\ee
When we add all the equations from (\ref{A2-6}) to (\ref{A2-15}), we find that (\ref{4.3}) holds.  This serves as a check on the algebra.

\section{Five-point $D$'s}\label{Appendix3} 

Following the procedure from (\ref{4.3}) to (\ref{4.8}), we
arrive at the other $D$'s:
\bea
D(14325)&=& \Delta (14|352)({1\over s_{14}s_{23}}
+{1\over s_{23}s_{15}})
+\Delta (12|345)(-{1\over s_{34}s_{15}})\nonumber\\ && 
+\Delta (25|134)(-{1\over s_{34}s_{15}}
-{1\over s_{25}s_{34}})
+\Delta (34|125)(-{1\over s_{34}s_{15}}) \nonumber\\ &&
+\Delta (45|123)({1\over s_{23}s_{15}})
+\Delta (23|145)({1\over s_{23}s_{15}}), \label{A3-1}
\\
D(13425)&=& \Delta (13|245)({1\over s_{13}s_{24}}
+{1\over s_{24}s_{15}})
+\Delta (12|345)({1\over s_{34}s_{15}})\nonumber\\ && 
+\Delta (25|134)({1\over s_{34}s_{15}}
+{1\over s_{25}s_{34}})
+\Delta (34|125)({1\over s_{34}s_{15}}) \nonumber\\ &&
+\Delta (35|124)({1\over s_{24}s_{15}})
+\Delta (24/135)({1\over s_{24}s_{15}}),  \label{A3-2}
\\
D(12435)&=& \Delta (12|345)({1\over s_{12}s_{34}}
+{1\over s_{34}s_{15}})
+\Delta (24|135)({1\over s_{24}s_{15}})\nonumber\\ && 
+\Delta (35|124)({1\over s_{24}s_{15}}
+{1\over s_{24}s_{35}})
+\Delta (13|245)({1\over s_{24}s_{15}}) \nonumber\\ &&
+\Delta (34|125)({1\over s_{34}s_{15}})
+\Delta (25|134)({1\over s_{34}s_{15}}),  \label{A3-3}
\\
D(14235)&=& \Delta (14|352)(-{1\over s_{14}s_{23}}
-{1\over s_{23}s_{15}})
+\Delta (24|135)(-{1\over s_{24}s_{15}})\nonumber\\ && 
+\Delta (35|124)(-{1\over s_{24}s_{15}}
-{1\over s_{24}s_{35}})
+\Delta (13|245)(-{1\over s_{24}s_{15}}) \nonumber\\ &&
+\Delta (45|123)(-{1\over s_{23}s_{15}})
+\Delta (23|145)(-{1\over s_{23}s_{15}}),  \label{A3-4}
\\
D(13245)&=& \Delta (13|245)(-{1\over s_{13}s_{24}}
-{1\over s_{24}s_{15}})
+\Delta (14|352)(-{1\over s_{23}s_{15}})\nonumber\\ && 
+\Delta (45|123)(-{1\over s_{23}s_{15}}
-{1\over s_{23}s_{45}})
+\Delta (23|145)(-{1\over s_{23}s_{15}}) \nonumber\\ &&
+\Delta (35|124)(-{1\over s_{24}s_{15}})
+\Delta (24|135)(-{1\over s_{24}s_{15}}). \label{A3-5}
\eea

\section{Solving for the five-point numerator shifts}\label{Appendix4}

In this appendix we give the details of the steps taken to arrive at the solution given in the main text for the numerator shifts.
 First we notice that
because of 
\be
a_6=a_1(2 \leftrightarrow 4)=-a_1(1\leftrightarrow 5), 
\label{4.24}
\ee
we have 
\bea
&&z_1^{12}(1\leftrightarrow 5)=-x_1^{54}(2 \leftrightarrow 4),
z_1^{14}(1\leftrightarrow 5)=-x_1^{52}(2 \leftrightarrow 4),\nonumber\\ &&
z_1^{32}(1\leftrightarrow 5)=-x_1^{34}(2 \leftrightarrow 4),
z_1^{34}(1\leftrightarrow 5)=-x_1^{32}(2 \leftrightarrow 4),\nonumber\\ &&
y_1^{12}(1\leftrightarrow 5)=-y_1^{54}(2 \leftrightarrow 4), 
y_1^{14}(1\leftrightarrow 5)=-y_1^{52}(2 \leftrightarrow 4). \label{4.25}
\eea
Instead of using (\ref{4.17}) or (\ref{4.18}) to normalize the $x$, $y$, $z$'s 
we use instead equivalently\footnote{We are using here notation introduced earlier in eqn. (3.1) in \cite{VY3}.} 
\be
\delta n_1 -\delta n_3 - \delta n_{12} =-\Delta(12|345), \label{4.26}
\ee
where 
\be
\delta n_3=\delta n(12;5;43)=\delta n_1(3\leftrightarrow 5).  \label{4.27}
\ee

We now use (\ref{4.21}) to obtain four independent equations:
\be
x_1^{32}+x_1^{34}(2 \leftrightarrow 4)
+x_{15}^{32} +x_{15}^{34}(2\leftrightarrow 4)
-z_{15}^{34}(2\leftrightarrow 4;1\leftrightarrow 5)
-z_{15}^{32}(1\leftrightarrow 5)=0, \label{4.28}
\ee

\bea
&&x_1^{52}+x_1^{54}(2 \leftrightarrow 4)
+x_{15}^{52} +x_{15}^{54}(2\leftrightarrow 4)
-z_{15}^{14}(2\leftrightarrow 4;1\leftrightarrow 5)
-z_{15}^{12}(1\leftrightarrow 5)=0, \label{4.29}\\
&&
y_1^{12}+y_1^{14}(2 \leftrightarrow 4)
+y_{15}^{12} +y_{15}^{14}(2\leftrightarrow 4)
-y_{15}^{54}(2\leftrightarrow 4;1\leftrightarrow 5)
-y_{15}^{52}(1\leftrightarrow 5)=0, \label{4.30}\\
&&
y_1^{52}+y_1^{54}(2 \leftrightarrow 4)
+y_{15}^{52} +y_{15}^{54}(2\leftrightarrow 4)
-y_{15}^{14}(2\leftrightarrow 4;1\leftrightarrow 5)
-y_{15}^{12}(1\leftrightarrow 5)=0. \label{4.31}
\eea
By equating coefficients multiplied to different ${p_i^-\over p_i}
p_{jk}$ from (\ref{4.26}), we obtain a set of twelve equations:
\bea
&&-s_{34}x_1^{52}(3\leftrightarrow 5)+s_{45}x_1^{32}
+s_{35}z_{15}^{34}(2\leftrightarrow 4; 1\leftrightarrow 5)=1,
\label{4.32},\\
&&-s_{34}x_1^{54}(3\leftrightarrow 5)+s_{45}x_1^{34}
+s_{35}z_{15}^{32}(2\leftrightarrow 4; 1\leftrightarrow 5)=0,
\label{4.33}\\
&&-s_{34}x_1^{32}(3\leftrightarrow 5)+s_{45}x_1^{52}
+s_{35}z_{15}^{14}(2\leftrightarrow 4; 1\leftrightarrow 5)=-1,
\label{4.34}\\
&&
-s_{34}x_1^{34}(3\leftrightarrow 5)+s_{45}x_1^{54}
+s_{35}z_{15}^{12}(2\leftrightarrow 4; 1\leftrightarrow 5)=0;
\label{4.35}\\
&&s_{34}x_1^{54}(2\leftrightarrow 4;1\to 3\to 5 \to 1)
+s_{45}y_1^{12}+s_{35}y_{15}^{54}(2\leftrightarrow 4;1\leftrightarrow 5)=0,
\label{4.36}\\
&&s_{34}x_1^{52}(2\leftrightarrow 4;1\to 3\to 5 \to 1)
+s_{45}y_1^{14}+s_{35}y_{15}^{52}(2\leftrightarrow 4;1\leftrightarrow 5)=-1,
\label{4.37}\\
&&s_{34}x_1^{34}(2\leftrightarrow 4;1\to 3\to 5 \to 1)
+s_{45}y_1^{52}+s_{35}y_{15}^{14}(2\leftrightarrow 4;1\leftrightarrow 5)=1,
\label{4.38}\\
&&s_{34}x_1^{32}(2\leftrightarrow 4;1\to 3\to 5 \to 1)
+s_{45}y_1^{54}+s_{35}y_{15}^{12}(2\leftrightarrow 4;1\leftrightarrow 5)=0;
\label{4.39}\\
&&-s_{34}y_1^{12}(3\leftrightarrow 5)+s_{45}z_1^{12}
+s_{35}x_{15}^{54}(2\leftrightarrow 4; 1\leftrightarrow 5)=0, \label{4.40}\\
&&-s_{34}y_1^{14}(3\leftrightarrow 5)+s_{45}z_1^{14}
+s_{35}x_{15}^{52}(2\leftrightarrow 4; 1\leftrightarrow 5)=1, \label{4.41}\\
&&-s_{34}y_1^{52}(3\leftrightarrow 5)+s_{45}z_1^{32}
+s_{35}x_{15}^{34}(2\leftrightarrow 4; 1\leftrightarrow 5)=-1, \label{4.42}\\
&&-s_{34}y_1^{54}(3\leftrightarrow 5)+s_{45}z_1^{34}
+s_{35}x_{15}^{32}(2\leftrightarrow 4; 1\leftrightarrow 5)=0. \label{4.43}
\eea
When we make $3\leftrightarrow 5$ to (\ref{4.34}), we obtain
\be
-s_{45}x_1^{32}+s_{34}x_1^{52}+s_{35}
z_{15}^{14}(2\leftrightarrow 4;1\to 3\to 5\to 1)=-1, \label{4.34'}
\ee
which is added to (\ref{4.32}) to give 
\be
z_{15}^{32}(1\leftrightarrow 5)+z_{15}^{12}(1\to 3\to 5\to 1)=0. \label{4.44}.
\ee
In a similar fashion, we obtain from (\ref{4.33}) and (\ref{4.35})
\be
z_{15}^{34}(1\leftrightarrow 5)+z_{15}^{14}(1\to 3\to 5\to 1)=0. \label{4.45}.
\ee
If we use the results above, then we should of course keep only one of 
(\ref{4.32}) and (\ref{4.34}) and one of (\ref{4.33}) and (\ref{4.35}).

Using $z_1^{12}=-x_1^{54}(2\leftrightarrow 4; 1\leftrightarrow 5)$ 
and making $3\leftrightarrow 5$, we write (\ref{4.40}) as
\be
-s_{45}y_1^{12}-s_{34}x_1^{54}(2\leftrightarrow 4; 1\to 3\to 5\to 1)
+s_{35}x_{15}^{54}(2\leftrightarrow 4; 1\to 3\to 5\to 1)=0. \label{4.40'}
\ee
When we combine this with (\ref{4.36}). we have
\be
x_{15}^{54}(1\leftrightarrow 3)+y_{15}^{54}=0.  \label{4.46}
\ee
The same operations will lead to 
\bea
&&x_{15}^{52}(1\leftrightarrow 3)+y_{15}^{52}=0.  \label{4.47}\\
&&x_{15}^{34}(1\leftrightarrow 3)+y_{15}^{14}=0.  \label{4.48}\\
&&x_{15}^{32}(1\leftrightarrow 3)+y_{15}^{32}=0.  \label{4.49}
\eea
We should then keep either the set (\ref{4.36}) to (\ref{4.39}) or the set 
(\ref{4.40}-\ref{4.43}).
Therefore we have only six  of equations (\ref{4.32}) to (\ref{4.43}) and the four 
of equations (\ref{4.28}) to (\ref{4.31}), which add up to ten.  
Taking into account (\ref{4.25})
and (\ref{4.44}) to (\ref{4.49}), we have twelve independent equations for $x$'s, $y$'s and $z$'s. 
We noted earlier that these coefficients have dimension $1/s$. We will be solving for them with the requirement that they are of the form of a sum of terms each being a simple pole in the allowed kinematic invariant (such that the numerator shifts do not introduce spurious poles). This  leads to the solution given in (\ref{4.50}) and (\ref{4.51}).

\section{Beyond five-point}\label{Appendix5}

In this appendix we discuss how one can extend recursively the current results beyond five-points. 
 Consider the six-point case. We begin by choosing a Kleiss-Kuijf basis  as in \cite{VY3}: $A(1 i_2 i_3 i_4 i_5 6)$ with $(i_2,i_3,i_4,i_5)$ equal to a permutation of indices (2,3,4,5). 
We use the shorthand notation\footnote{This type of numerators has been later called half-ladder in \cite{Broedel:2011pd}. The reason is that the external legs are all arranged along an internal line with two external legs joined together only at the two ends of that internal line.} 
\bea
&&n(12;3;4;56)=n_1,\qquad n(13; 2;4;56)=n_2,\qquad n(13;4;2;56)=n_3,\qquad
n(13;4;5;26)=n_4\nn\\
&&n(12;4;3;56)=n_5,\qquad n(14;2;3;56)=n_6,\qquad n(14;3;2;56)=n_7,\qquad
n(14;3;5;26)=n_8\nn\\
&&n(12;5;4;36)=n_9,\qquad n(15;2;4;36)=n_{10},\qquad n(15;4;2;36)=n_{11},\qquad
n(15;4;3;26)=n_{12}\nn\\
&&n(12;3;5;46)=n_{13},\qquad n(13;2;5;46)=n_{14},\qquad n(13;5;2;46)=n_{15},\qquad n(13;5;4;26)=n_{16}
\nn\\
&&n(12;4;5;36)=n_{17},\qquad n(14;2;5;36)=n_{18},\qquad n(14;5;2;36)=n_{19},\qquad n(14;5;3;26)=n_{20}\nn\\
&&n(12;5;3;46)=n_{21},\qquad n(15;2;3;46)=n_{22},\qquad n(15;3;2;46)=n_{23},\qquad n(15;3;4;26)=n_{24}.\nn\\
\label{KK6}
\eea

For each color-ordered amplitude we decompose into terms which display the propagator pole structure as in eqn (A.2) in \cite{VY3}. For example,
\bea
A(123456)&=&\frac{n_1}{s_{12}s_{123}s_{1234}}-
\frac{n(12;3;6;45)}{s_{12}s_{123}s_{1236}}-
\frac{n(12;6;3;45)}{s_{12}s_{126}s_{1236}}\nn\\
&+&\frac{n(61;2;3;45)}{s_{16}s_{126}s_{1236}}+
\frac{n(12;6;5;34)}{s_{12}s_{126}s_{1256}}+
\frac{n(23;4;5;61)}{s_{23}s_{234}s_{2345}}\nn\\
&-&\frac{n(23;4;1;56)}{s_{23}s_{234}s_{1234}}-
\frac{n(23;1;4;56)}{s_{23}s_{123}s_{1234}}
+\frac{n(34;2;1;56)}{s_{34}s_{234}s_{1234}}\nn\\
&-&\frac{n(34;5;2;61)}{s_{34}s_{345}s_{2345}}
-\frac{n(34;2;5;61)}{s_{34}s_{234}s_{2345}}
+\frac{n(23;1;6;45)}{s_{23}s_{123}s_{1236}}\nn\\
&+&\frac{n(12;34;56)}{s_{12}s_{34}s_{56}}
+\frac{n(61;23;45)}{s_{16}s_{23}s_{45}}.
\eea
However, the Feynman-rules numerators will not satisfy the BCJ relations (as opposed to the numerators in eqn (A.5) of \cite{VY3}). Instead, there will be violations which we parametrized as in Appendix \ref{Appendix2} by $\Delta$'s. These can be constructed as follows.
 For concreteness let us focus on
\be
n(12;3;4;56)+n(12;3;6;45)+n(12;3;5;64)=\Delta(12;3|456) ,
\ee
where
\bea
\Delta(12;3|456)&=&(12k) \Delta(k3|456)
                        +(12;3k)\Delta(k|456) .
\eea
and as before $(12k)$ denotes a three-point vertex and $(12;3k)$ denotes a four-point vertex\footnote{Recall that according to the Feynman rules the four-point vertex contribution $(12;3k)$ is non-zero only when the gluons in each pair (12) and $(3k)$ have opposite helicities, and that $(12;3k)$ is proportional to $s_{12}$.}. The off-shell five-point $\Delta$'s are given by the corresponding version of (\ref{A2-4}) plus off-shell terms. For example $\Delta(1^+2^-|3^+4^-5^+)$, where all legs are taken to be off-shell, has the following off-shell pieces (representing the contributions of the off-shell four-point $\Delta$ to\footnote{Recall that $\Delta(k^+|3^+4^-5^+)=0$ and that 
$\Delta(k|345)=\Delta(543|k)=\Delta(435|k)$.} $(1^+2^-k^+)\Delta(k^-|3^+4^-5^+)$) in addition to (\ref{A2-4}): 
\bea
&&\bigg[\frac{p_3 (p_1+p_2)}{p_3+p_4}\bigg(-\frac{P_5^2}{2p_5}+\frac{P_4^2}{2p_4}\bigg) 
+\frac{p_4 p_5}{p_4+p_3}\bigg(-\frac{(P_1+P_2)^2}{2(p_1+p_2)}+\frac{P_3^2}{2p_3}\bigg)\nn\\
&&+\frac{p_5(p_1+p_2)}{p_4+p_5}\bigg(-\frac{P_3^2}{2p_3}+\frac{P_4^2}{2p_4}\bigg)
-\frac{p_4 p_3}{p_4+p_5}\bigg(-\frac{(P_1+P_2)^2}{2(p_1+p_2)}+\frac{P_5^2}{2p_5}\bigg)\bigg]\nn\\
&&\times(\frac{p_1^-+p_2^-}{p_1+p_2}-\frac{p_1^-}{p_1})p_2.
\eea

Consider $n(12;6;3;45)$ as obtained by Feynman rules. We can write this as 
$(12k)n(k6;3;45)+ (12;6k)n(k3;45)$. The second term is necessary since it is a contribution from the 4-point vertex $(12;6k)$ which is not included in the first term where a cubic vertex is affixed to the off-shell 5-point numerator.

Next we use that 
\be
n(k6;3;45)+n(3k;6;45)+n(63;k;45)=\Delta(k63|45)
\ee
 where this is the off-shell 5-point $\Delta$ described earlier in this section.

Then 
\be
n(3k;6;45)=n(k3;4;56)+n(k3;5;64)+\Delta(3k|645),
\ee
 while
\be
n(63;k;45)=n(36;4;5k)+n(36;5;k4)+\Delta(63|k45).
\ee 

Putting everything together, 
\bea
n(12;6;3;45)&=&(12k)n(k6;3;45)+ (12;6k)n(k3;45)\nn\\
&=&(12k)[\Delta(k63|45)+\Delta(k3|645)-\Delta(63|k45)]+(12;6k)n(k3;45)\nn\\
&-&(12k)n(k3;4;56)-[(12;3k)n(k456)-(12;3k)n(k4;56)]\nn\\
&+&(12k)n(k3;5;46)+[(12;3k)n(k5;46)-(12;3k)n(k5;46)]\nn\\
&+&(12k)n(k5;4;63)+[(12;5k)n(k4;63)-(12;5k)n(k4;63)]\nn\\
&+&(12k)n(k4;5;36)+[(12;4k)n(k5;36)-(12;4k)n(k5;36)]\nn\\
&=&(12k)[\Delta(k63|45)+\Delta(k3|645)-\Delta(63|k45)]+(12;6k)n(k3;45)\nn\\
&-&n(12;3;4;56)+n(12;3;5;46)-n(12;5;4;36)+n(12;4;5;36)\nn\\
&+&(12;3k)n(k4;56)-(12;3k)n(k5;46)+(12;5k)n(k4;36)-(12;4k)n(k5;36)\nn\\
%
&=&n(12;3;4;56)+n(12;3;5;46)-n(12;5;4;36)+n(12;4;5;36)\nn\\
&+&\Delta(12;3|645)\nn\\
&+&(12k)\Delta(36|k45)+(12;6k)n(k3;45)-(12;3k)n(k6;45)\nn\\
&+&(12k)\Delta(45|36k)+(12;5k)n(k4;36)-(12;4k)n(k5;36).
\eea

The following numerators can be expressed in this way and obtained by relabelling of external legs:
\be
n(12;6;5;34)\qquad{\rm {from}}\qquad n(12;6;3;45)\qquad{\rm {with}}\qquad (3\to5,4\to3,5\to4)
\ee
\be
n(23;1;4;56)=n(65;4;1;32) \qquad{\rm {from}} \qquad n(12;6;3;45) \qquad{\rm {with}}\qquad (1\leftrightarrow 6,5\leftrightarrow 2, 4\leftrightarrow 3)
\ee
\be
n(34;2;1;56)=n(65;1;2;43) \qquad{\rm {from}}\qquad
n(12;6;3;45)  \qquad{\rm {with}}\qquad
(6\leftrightarrow1, 5\to2,2\to3,3\to5).
\ee

Yet another type of terms is $n(23;1;6;45)$. We write it as 
$n(23;1k)n(k6;45)+(23k)(k1;6l)(l45)$ and manipulate it such that we express it in terms of the chosen basis of numerators plus violating terms.
\bea
n(23;1;6;45)&=&(-n(12;3k)-n(31;2k)+\Delta(123|k))
(-n(k4;56)-n(k5;64)+\Delta(k|645))\nn\\
&+&(23k)(k1;6l)(l45)\nn\\
&=&n(12;3;4;56)-n(12;3;5;46)-n(13;2;4;56)+n(13;2;5;46)\nn\\
&-&\Delta(231|k)[n(k4;56)+n(k5;64)]-\Delta(k|645)[n(12;3k)+n(31;2k)]\nn\\
&-&(12k)[(k3;4l)(l56)+(k3;5l)(l64)]+(13k)[(k2;4l)(l56)+(k2;5l)(l64)]\nn\\
&+&(23k)(k1;6l)(l45).
\eea

Then we have the snowflake $n(12;34;56)$. This can be expressed as 
$-(12k)n(34;k;56)+(34;k)(kl;12)(l56)+(12k)(34;lk)(l56)+(12k)(kl;56)(l34)$, so
\bea
n(12;34;56)&=&-(12k)\Delta(34k|56)+(12k)n(4k;3;56)+(12k)n(k3;4;56)\nn\\
&+&(34;k)(kl;12)(l56)+(12k)(34;lk)(l56)+(12k)(kl;56)(l34)\nn\\
&=&-(12k)\Delta(34k|56)-n(12;4;3;56)+n(12;3;4;56)\nn\\
&+&(12k)(k4;3l)(l56)-(12k)(k4;3l)(l56)\nn\\
&+&(34;k)(kl;12)(l56)+(12k)(34;lk)(l56)+(12k)(kl;56)(l34).\nn\\
\eea

The more complicated numerators have an $s_{61}$ associated pole.
Let's consider $n(61;2;3;45)$. We can write it as $n(61;2k)n(k3;45)+(61k)(k2;3l)(l45)$, which gives
\bea
n(61;2;3;45)&=&[-n(12;6k)-n(26;1k)+\Delta(162|k)]n(k3;45)+(61k)(k2;3l)(l45)\nn\\
&=&-n(12;6;3;45)+n(62;1;3;45)+\Delta(162|k)n(k3;45)\nn\\
&+&(61k)(k2;3l)(l45)+(12k)(k6;3l)(l45)+(26k)(k1;3l)](l45),
\eea
then each of the numerators $n(12;6;3;45)$ and $n(26;1;3;45)$ receives the same treatment as before.
For the final expression,
\bea
n(61;2;3;45)&=&n(12;3;4;56)-n(12;3;5;46)+n(12;5;4;36)-n(12;4;5;36)\nn\\
&-&n(62;3;4;51)+n(62;3;5;41)-n(62;5;4;31)+n(62;4;5;31)\nn\\
&-&\Delta(12;3|645)\nn\\
&-&(12k)\Delta(36|k45)-(12;6k)n(k3;45)-(12;k3)n(k6;45)\nn\\
&-&(12k)\Delta(45|36k)-(12;5k)n(k4;36)-(12;k4)n(k5;36)\nn\\
&+&\Delta(15;4|632)\nn\\
&+&(15k)\Delta(46|k32)+(15;6k)n(k4;32)-(15;k4)n(k6;32)\nn\\
&+&(15k)\Delta(32|46k)+(15;2k)n(k3;46)-(15;k3)n(k2;46)\nn\\
&+&\Delta(162|k)n(k3;45)\nn\\
&+&(61k)(k2;3l)(l45)+(12k)(k6;3l)(l45)+(26k)(k1;3l)(l45).
\eea
The other numerators in the same family are obtained as follows
\bea
n(23;4;5;61)&=&n(16;5;4;32)=-n(61;5;4;32)
\qquad{\rm {from}} -n(61;2;3;45) \qquad{\rm {with}}\qquad (2\leftrightarrow 5, 3\leftrightarrow 4)
\nn\\
n(34;5;2;61)&=&-n(61;2;5;43) \qquad{\rm {from}}\qquad-n(61;2;3;45) \qquad{\rm {with}}\qquad 
(3\leftrightarrow 5)
\nn\\
n(34;2;5;61)&=&-n(61;5;2;43) \qquad{\rm {from}}\qquad -n(61;2;3;45) \qquad{\rm {with}}\qquad 
(2\to5, 3\to2,5\to3).\nn\\
\eea

Lastly, we have the snowflake with a $s_{61}$ inverse propagator. We write it as
\be
n(61;23;45)=n(61kl)(k23)(l45)+(16k)(23;kl)(l45)+(16k)(45;kl)(l23),
\ee
to obtain
\bea
n(61;23;45)&=&-(23k)n(k1;6l)(l45)+(23k)n(k6;1l)(l45)\nn\\&+&
\Delta(61k|l)(k23)(l45)+(16k)(23;kl)(l45)+(16k)(45;kl)(l23)
\nn\\
&=&-n(23;1;6;45)+n(23;6;1;45)\nn\\
&+&(23;1k)n(k6;45)+n(23;1k)(k6;45)
-(23;6k)n(k1;45)-n(23;6k)(k1;45)\nn\\
&+&
\Delta(61k|l)(k23)(l45)+(16k)(23;kl)(l45)+(16k)(45;kl)(l23).
\eea
The numerators $n(23;1;6;45)$ and $n(23;6;1;45)$ have been discussed before,
leading to the following expression for the 61-snowflake:
\bea
n(61;23;45)&=&-\bigg( n(12;3;4;56)-n(12;3;5;46)-n(13;2;4;56)+n(13;2;5;46)\nn\\
&-&\Delta(231|k)[n(k4;56)+n(k5;64)]-\Delta(k|645)[n(12;3k)+n(31;2k)]\nn\\
&-&(12k)[(k3;4l)(l56)+(k3;5l)(l64)]+(13k)[(k2;4l)(l56)+(k2;5l)(l64)]\nn\\
&+&(23k)(k1;6l)(l45)\bigg)
\nn\\
&+&\bigg( n(62;3;4;51)-n(62;3;5;41)-n(63;2;4;51)+n(63;2;5;41)\nn\\
&-&\Delta(236|k)[n(k4;51)+n(k5;64)]-\Delta(k|145)[n(62;3k)+n(36;2k)]\nn\\
&-&(62k)[(k3;4l)(l51)+(k3;5l)(l14)]+(63k)[(k2;4l)(l51)+(k2;5l)(l14)]\nn\\
&+&(23k)(k6;1l)(l45)
\bigg)\nn\\
&+&(23;1k)n(k6;45)+n(23;1k)(k6;45)
-(23;6k)n(k1;45)-n(23;6k)(k1;45)\nn\\
&+&
\Delta(61k|l)(k23)(l45)+(16k)(23;kl)(l45)+(16k)(45;kl)(l23).
\eea

Armed with this we can proceed to computing the $D$'s. For example, by collecting together all the $\delta n$-independent terms in the expression below gives $D(123456)$, in a natural extension of the five-point relations (\ref{4.3}-\ref{4.8}):
\bea
&&\frac{\delta n_1}{s_{12} s_{123} s_{1234}}
- \frac{-\delta n_1+\delta n_{13} + \Delta(12;3|456)}{s_{12} s_{123}s_{1236}}
+\frac{-\delta n_1+\delta n_2 + \Delta(231|4;56)}{s_{23}s_{123}s_{1234}}\nn\\
&&-\frac{1}{s_{12} s_{126} s_{1236}}\bigg(
-\delta n_1 - \delta n_9 + \delta n_{13} + \delta n_{17} +\Delta(12;3|645)\nn\\
&&+(12k)\Delta(36|k45)+(12;6k)n(k3;45)+(12;3k)n(6k;45)\nn\\
&&+(12k)\Delta(45|36k)+(12;5k)n(k4;36)+(12;k4)n(k5;36)\bigg)
\nn\\
&&+\frac{1}{s_{12} s_{126} s_{1256}}\bigg(
\delta n_1 - \delta n_5 + \delta n_{9} - \delta n_{21} 
+\Delta(12;5|634)\nn\\
&&+(12k)\Delta(56|k34)+(12;6k)n(k5;34)+(12;5k)n(6k;34)\nn\\
&&+(12k)\Delta(34|56k)+(12;4k)n(k3;56)+(12;k3)n(k4;56)\bigg)\nn\\
&&-\frac1{s_{23}s_{243}s_{1234}}\bigg(
-\delta n_1+\delta n_2+\delta n_6-\delta n_7+
+\Delta(65;4|132)\nn\\
&&+(65k)\Delta(41|k32)+(65;1k)n(k4;32)+(65;4k)n(1k;32)\nn\\
&&+(65k)\Delta(32|41k)+(65;2k)n(k3;41)+(65;k3)n(k2;41)\bigg)
\nn\\
&&+\frac{1}{s_{34}s_{234}s_{1234}}
\bigg(\delta n_1-\delta n_3-\delta n_5+\delta n_7
+\Delta(62;4|135)\nn\\
&&+(64k)\Delta(31|k52)+(64;1k)n(k3;52)+(64;3k)n(1k;52)\nn\\
&&+(64k)\Delta(52|31k)+(64;2k)n(k5;31)+(64;k5)n(k2;41)\bigg)\nn\\
&&+\frac1{s_{23}s_{123}s_{1236}}
\bigg(\delta n_1-\delta n_2-\delta n_{13}+\delta n_{14}\nn\\&&
-\Delta(231|k)[n(k4;56)+n(k5;64)]-\Delta(k|645)[n(12;3k)+n(31;2k)]\nn\\
&&-(12k)[(k3;4l)(l56)+(k3;5l)(l64)]+(13k)[(k2;4l)(l56)+(k2;5l)(l64)]\nn\\
&&+(23k)(k1;6l)(l45)\bigg)\nn\\
&&+\frac1{s_{12}s_{34}s_{56}}\bigg(\delta n_1-\delta n_5 
-(12k)\Delta(34k|56)\nn\\
&&+(12k)(k4;3l)(l56)-(12k)(k4;3l)(l56)\nn\\
&&+(34;k)(kl;12)(l56)+(12k)(34;lk)(l56)+(12k)(kl;56)(l34)\bigg)\nn
\eea
\bea
&&+\frac{1}{s_{16}s_{126}s_{1236}}
\bigg( 
\delta n_1-\delta n_4+\delta n_9-\delta n_{12}-\delta n_{13}+\delta n_{16}-\delta n_{17}+\delta n_{20}\nn\\
&&-\Delta(12;3|645)+\Delta(15;4|632)
-(12k)\Delta(36|k45)-(12;6k)n(k3;45)-(12;k3)n(k6;45)\nn\\
&&-(12k)\Delta(45|36k)-(12;5k)n(k4;36)-(12;k4)n(k5;36)
+(15k)\Delta(46|k32)+(15;6k)n(k4;32)\nn\\
&&-(15;k4)n(k6;32)
+(15k)\Delta(32|46k)+(15;2k)n(k3;46)-(15;k3)n(k2;46)\nn\\
&&+\Delta(162|k)n(k3;45)
+(61k)(k2;3l)(l45)+(12k)(k6;3l)(l45)+(26k)(k1;3l)(l45)\bigg)
\nn\\
&&-\frac1{s_{16}s_{156}s_{1456}}
\bigg(\delta n_1-\delta n_2-\delta n_6+\delta n_7+\delta n_{11}-\delta n_{12}-\delta n_{22}+\delta n_{23}
\nn\\
&&-\Delta(15;4|632)+\Delta(12;3|645)
-(15k)\Delta(46|k32)-(15;6k)n(k4;32)-(15;k4)n(k6;32)\nn\\
&&-(15k)\Delta(32|46k)-(15;2k)n(k3;46)-(15;k3)n(k2;46)
+(12k)\Delta(36|k45)+(12;6k)n(k3;45)\nn\\
&&-(12;k3)n(k6;45)
+(12k)\Delta(45|36k)+(12;5k)n(k4;36)-(12;k4)n(k5;36)\nn\\
&&+\Delta(165|k)n(k4;32)
+(61k)(k5;4l)(l32)+(15k)(k6;4l)(l32)+(56k)(k1;4l)(l32)\bigg)
\nn\\
&&-\frac{1}{s_{16}s_{156}s_{1256}}
\bigg(-\delta n_1+\delta n_3+\delta n_5-\delta n_7-\delta n_{10}+\delta n_{12}+\delta n_{22}-\delta n_{24}\nn\\
&&-\Delta(15;2|643)+\Delta(13;4|625)
-(15k)\Delta(26|k43)-(15;6k)n(k2;43)-(15;k2)n(k6;43)\nn\\
&&-(15k)\Delta(43|26k)-(15;3k)n(k4;26)-(15;k4)n(k3;26)
+(13k)\Delta(46|k25)+(13;6k)n(k4;25)\nn\\
&&-(13;k4)n(k6;25)
+(13k)\Delta(25|46k)+(13;5k)n(k2;46)-(13;k2)n(k5;46)\nn\\
&&+\Delta(165|k)n(k2;43)
+(61k)(k5;2l)(l43)+(15k)(k6;2l)(l43)+(56k)(k1;2l)(l43)\bigg)
\nn\\
&&-\frac{1}{s_{16}s_{126}s_{1256}}\bigg(-\delta n_1+\delta n_4+\delta n_5-\delta n_8-\delta n_{9 }+\delta n_{12}+\delta n_{21}-\delta n_{24}\nn\\
&&-\Delta(12;3|645)+\Delta(15;4|632)
-(12k)\Delta(56|k43)-(12;6k)n(k5;43)-(12;k5)n(k6;43)\nn\\
&&-(12k)\Delta(43|56k)-(12;3k)n(k4;56)-(12;k4)n(k3;56)
+(13k)\Delta(46|k52)+(13;6k)n(k4;52)\nn\\
&&-(13;k4)n(k6;52)
+(13k)\Delta(52|46k)+(13;2k)n(k5;46)-(13;k5)n(k2;46)\nn\\
&&+\Delta(162|k)n(k5;43)
+(61k)(k2;5l)(l43)+(12k)(k6;5l)(l43)+(26k)(k1;5l)(l43)\bigg)
\nn
\\
&&+\frac{1}{s_{16}s_{23}s_{45}}\bigg(\delta n_1-\delta n_2+\delta n_{11}-\delta n_{12}-\delta n_{13}+\delta n_{14}-\delta n_{19}+\delta n_{20}\nn\\
&&+\Delta(231|k)[n(k4;56)+n(k5;64)]+\Delta(k|645)[n(12;3k)+n(31;2k)]\nn\\
&&+(12k)[(k3;4l)(l56)+(k3;5l)(l64)]-(13k)[(k2;4l)(l56)+(k2;5l)(l64)]
-(23k)(k1;6l)(l45)
\nn\\
&&-\Delta(236|k)[n(k4;51)+n(k5;64)]-\Delta(k|145)[n(62;3k)+n(36;2k)]\nn\\
&&-(62k)[(k3;4l)(l51)+(k3;5l)(l14)]+(63k)[(k2;4l)(l51)+(k2;5l)(l14)]
+(23k)(k6;1l)(l45)\nn\\
&&+(23;1k)n(k6;45)+n(23;1k)(k6;45)
-(23;6k)n(k1;45)-n(23;6k)(k1;45)\nn\\
&&+
\Delta(61k|l)(k23)(l45)+(16k)(23;kl)(l45)+(16k)(45;kl)(l23)
\bigg)\nn\\
&&=0\label{long}
\eea


\end{document}